\documentclass[hyper,letterpaper,12pt]{article}
\usepackage{amsmath}
\numberwithin{equation}{section}
\usepackage[
      colorlinks=true,
      linkcolor=blue,
      urlcolor=blue,
      filecolor=black,
      citecolor=blue,
            ]{hyperref}
\usepackage{color}
\usepackage{graphicx}
\usepackage{amsfonts}
\usepackage{amssymb}
\usepackage{hyperref}
\usepackage{simplewick}
\usepackage{breqn}
\usepackage{autobreak}
\usepackage{dsfont}

\graphicspath{{./figures/}}
\allowdisplaybreaks[4]

\begin{document}

\title{\bf \Large Color-Kinematics Duality for Sudakov Form Factor in Non-Supersymmetric Pure Yang-Mills Theory}

\author
{\large
~Zeyu Li\footnote{E-mail: lizeyu@itp.ac.cn}~,
~~Gang Yang\footnote{E-mail: yangg@itp.ac.cn}~,
~~Jinxuan Zhang\footnote{E-mail: zhangjinxuan@itp.ac.cn}\\
\\ 
\small \emph{$^1$CAS Key Laboratory of Theoretical Physics, Institute of Theoretical Physics,}\\
\small \emph{Chinese Academy of Sciences, Beijing 100190, China}\\
\small \emph{$^2$School of Physical Sciences, University of Chinese Academy of Sciences,} \\
\small \emph{No. 19A Yuquan Road, Beijing 100049, China}
}
\date{}
\maketitle

\begin{abstract}
\normalsize 
\noindent
We study the duality between color and kinematics for the Sudakov form factors of ${\rm tr}(F^2)$ in non-supersymmetric pure Yang-Mills theory. We construct the integrands that manifest the color-kinematics duality up to two loops. The resulting numerators are given in terms of Lorentz products of momenta and polarization vectors, which have the same powers of loop momenta as that from the Feynman rules. The integrands are checked by $d$-dimensional unitarity cuts and are valid in any dimension. We find that massless-bubble and tadpole topologies are needed at two loops to realize the color-kinematics duality. Interestingly, the two-loop solution contains a large number of free parameters suggesting the duality may hold at higher loop orders.

\end{abstract}

\newpage
\tableofcontents

\section{Introduction}

Significant progress has been made in the study of scattering amplitudes in the past thirty years, see \emph{e.g.~}\cite{Mangano:1990by, Dixon:1996wi, Bern:2007dw, 2011JPhA44a0101R, Dixon:2013uaa, Elvang:2013cua, Henn:2014yza, Feng:2011np, Weinzierl:2016bus, Alday:2008yw} for review. 
These studies have not only important phenomenological applications but also have uncovered various new structures that are not obvious at all from the traditional Feynman diagram method.
In this paper we focus on one of such structures, the so-called color-kinematics (CK) duality discovered by Bern, Carrasco, and Johansson \cite{Bern:2008qj, Bern:2010ue}.
The duality conjectures that there exists a trivalent Feynman-like diagrammatic representation of amplitudes in which the kinematic numerators satisfy the same algebraic relations as the color factors associated with the same graphs. Importantly, this duality indicates a deep connection between the gauge and gravity theories: via double copy  \cite{Bern:2010ue, Bern:2010yg}, one can obtain gravitational amplitudes directly from gauge amplitudes once the latter are organized to respect the CK duality, which is also closely related to the KLT relations \cite{Kawai:1985xq} and the CHY formula \cite{Cachazo:2013hca, Cachazo:2014xea}. 
See \cite{Bern:2019prr} for an extensive review of the duality and its applications.

Although the CK duality has been understood at tree level using \emph{e.g.}~string theory monodromy relations \cite{BjerrumBohr:2009rd, Stieberger:2009hq} or gauge theory recursion relations \cite{Feng:2010my}, it remains a conjecture at loop level.
In supersymmetric theories, the CK duality has been found to exist at high loop orders. For example, in the maximally supersymmetric ${\cal N}=4$ super-Yang-Mills (SYM) theory, the four-gluon amplitude has been found to preserve the duality up to four loops \cite{Bern:2010ue,Bern:2012uf} and for five-point amplitudes up to three loops \cite{Carrasco:2011mn}. For the form factors of the stress-tensor multiplet in ${\cal N}=4$ SYM, the CK-dual representation has been obtained up to five loops for the two-point Sudakov form factor \cite{Boels:2012ew,Yang:2016ear} and up to four loops for the three-point form factor \cite{Lin:2021kht,Lin:2021qol, Lin:2021lqo}. 
High-loop constructions for half-maximally supersymmetric theories were also studied in \cite{Bern:2012cd, Bern:2012gh,Bern:2013uka,Bern:2013qca, Bern:2014sna, Johansson:2017bfl}.
In the case of non-supersymmetric theories, some examples at one- and two-loop orders were found \cite{Boels:2013bi, Bern:2013yya, Bern:2015ooa, Mogull:2015adi, He:2017spx, He:2016mzd, Geyer:2017ela, Geyer:2019hnn, Edison:2020uzf}.

Despite this much progress, it is far from clear to which extent the duality holds at general loop orders. 
For example, for the four-point amplitude in ${\cal N}=4$ SYM, although a simple four-loop CK-dual solution has been obtained for a while \cite{Bern:2012uf}, it has proven difficult to construct the five-loop integrand which could manifest full CK duality \cite{Bern:2017yxu, Bern:2017ucb}.
In the non-supersymmetric gauge theories, the CK-dual representations of loop integrands are generally much harder to obtain. For example, for the four-gluon amplitude in pure YM theory, such a representation is so far only obtained for the case with identical four-dimensional external helicities at two loops \cite{Bern:2013yya}. For the two-loop five-gluon amplitudes in pure YM theory with identical helicities, it was found that the numerators with twelve powers of loop momenta have to be used to realize the duality, which is far beyond seven powers expected from Feynman diagrams \cite{Mogull:2015adi}. 

In this paper, we explore this duality in the non-supersymmetric pure Yang-Mills theory by considering Sudakov form factor of operator $\text{tr(}F^2\text{)}$ in the theory, which is defined as
\begin{equation}
	\label{eq:def-sudakovFF}
	\hat{\cal F}_{2}(1,2) = \int d^d x \, e^{-i q \cdot x} \langle g(p_1) g(p_2) |{\rm tr}(F^2)(x) | 0\rangle  \,,
\end{equation}
where $p_i$ are on-shell massless momenta of external gluons and $q=p_1+p_2$ is the off-shell momenta associated with the operator.
Sudakov form factor plays a central role in the study of IR divergences of gauge theories \cite{Mueller:1979ih,Collins:1980ih,Sen:1981sd,Magnea:1990zb}. 
In pure YM theory, the two-loop Sudakov form factor of ${\rm tr}(F^2)$ was firstly obtained to all orders in 
the dimensional regularization regulator $\epsilon= (4-d)/2$ using Feynman diagrams in \cite{Gehrmann:2005pd} (see also \cite{Harlander:2000mg}). 
In this work, we will reproduce this result using the CK duality and unitarity method, and we would like to stress that our main concern here is the structure of the form factor integrand.

We will start with constructing an ansatz of integrand by using CK duality, and then we solve the ansatz by applying the unitarity-cut method \cite{Bern:1994zx, Bern:1994cg, Britto:2004nc}.
Once the result satisfies a spanning set of cut constraints, it is a physically correct one.
In this way, we obtain the integrand solutions up to two loops which not only manifest all dual Jacobi relations but also satisfy all possible cuts.
Compared to the results in ${\cal N}=4$ SYM \cite{Boels:2012ew}, there are several important new features for the pure YM Sudakov form factor. 
First, the $d$-dimensional unitarity cuts are necessary to obtain the full results, and the resulting integrands also depend on the dimensional regulator $\epsilon= (4-d)/2$.
Second, massless-bubble and tadpole topologies are needed in order to satisfy CK duality at two-loop order.
In contrast, the two-loop ${\cal N}=4$ Sudakov form factor is much simpler, which contains only a planar-ladder and a cross-ladder topology.
Our pure YM results are given in terms of Lorentz products of momenta and polarization vectors, and they are valid in arbitrary dimensions. The CK-dual numerators have the same powers of loop momentum as expected from Feynman diagrams. Moreover, we find that the two-loop solution space still contains a large number of free parameters. These imply that the CK duality may promisingly hold for higher loop form factors in pure YM theory.

The rest of the paper is organized as follows. 
In Section~\ref{sec:CKansatz}, we construct the ansatz of the form factor integrands by imposing that they manifest the CK duality.
In Section~\ref{sec:unitaritycuts}, we apply unitarity cuts to solve for the coefficients in the ansatz and also study the constraints related to the tadpoles and massless bubbles. 
In Section~\ref{sec:integration}, we perform integral reductions and show that all parameters cancel after integral reductions. We also discuss the difference between the integrated results of the $d$-dimensional and 4-dimensional integrands.
A summary and outlook are given in Section~\ref{sec:discussion}. 
Several appendices provide some further details.
In Appendix~\ref{app:CKrelation}, we give the full set of two-loop dual Jacobi relations. 
The explicit expressions of the two-loop master numerators are given in Appendix~\ref{app:2loopnum}.
In Appendix~\ref{app:masterintegral}, we provide explicit expressions for the IBP master integrals and the master coefficients.

\section{CK duality and ansatz construction}\label{sec:CKansatz}

In this section we apply CK duality to construct the integrand ansatz for Sudakov form factor of ${\rm tr}(F^2)$ in pure YM theory. After giving a brief review of the duality, we will demonstrate how to generate the ansatz in the one-loop case step by step. Next we perform a similar construction at two loops.

\subsection{Review of color-kinematics duality}

\begin{figure}[t]
	\centerline{\includegraphics[height=2cm]{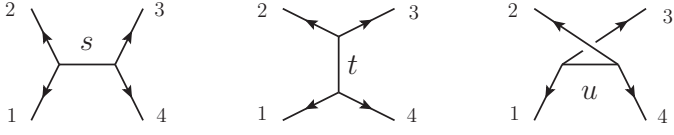} } 
	\caption{Trivalent graphs of the four-point tree amplitude.} 
	\label{fig:treeA4}
\end{figure}

In $SU(N_c)$ gauge theory, the structure constant is defined by
\begin{equation}\label{def:StrucCons}
	\tilde{f}^{abc}\text{=i}\sqrt{2}f^{abc}\text{=tr([}T^a,T^b\text{],}T^c\text{)}\,,
\end{equation}
in which $T^a$ are $SU(N_c)$ generators and are normalized by $\text{tr(}T^aT^b\text{)=}\delta ^{ab}$. 
By definition structure constants~\eqref{def:StrucCons} satisfy the Jacobi relation:
\begin{equation}\label{eq:Jacobi}
	\tilde{f}^{abe}\tilde{f}^{ecd}=\tilde{f}^{bce}\tilde{f}^{eda}+\tilde{f}^{ace}\tilde{f}^{ebd}\,.
\end{equation}
As depicted in Section 3 of~\cite{Bern:2008qj}, 
one can represent the full-color four-point tree amplitude as
\begin{equation}\label{eq:fullcolortreeA4}
	\mathcal{A}_4^{\rm tree}=g^2\left(\frac{c_sn_s}{s}+\frac{c_tn_t}{t}+\frac{c_un_u}{u}\right)\,,
\end{equation}
which correspond to $s, t, u$-channels shown in Figure~\ref{fig:treeA4}. 
The color factors satisfy the Jacobi relation 
\begin{equation}
	c_s=c_t+c_u \,.
\end{equation} 
It turns out that the numerators also satisfy a similar linear relation as: 
\begin{equation}\label{eq:treeDualJocabi}
	n_s=n_t+n_u\,.
\end{equation}
This correspondence is called color-kinematics (CK) duality. The relation \eqref{eq:treeDualJocabi} for the numerators is named as dual Jacobi relation or CK relation.

\begin{figure}[t]
	\centerline{\includegraphics[height=2.7cm]{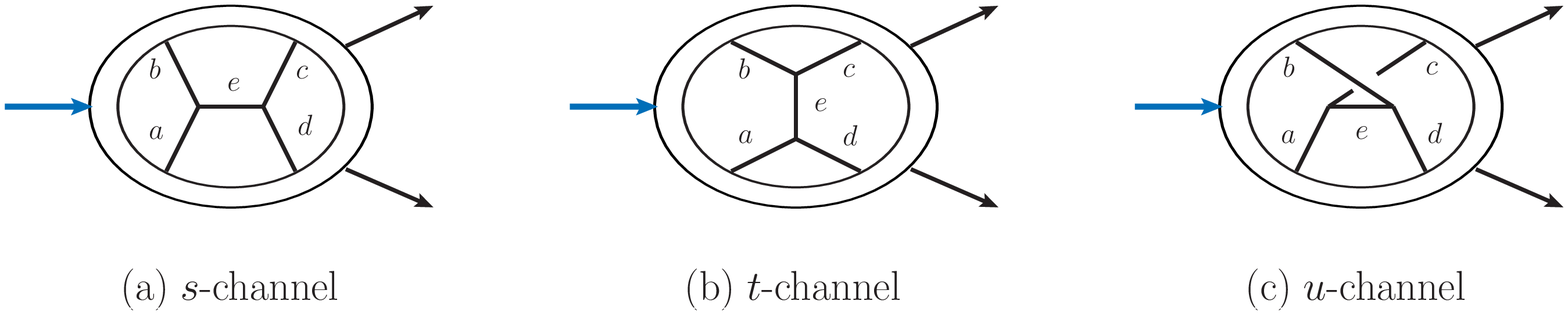} } 
	\caption{Loop diagrams related by Jacobi relation.} 
	\label{fig:diagramJacobi}
\end{figure}

One can impose similar color-kinematics duality at loop level as follows. For a certain loop diagram with color-dressed vertices, if we replace a four-point sub-diagram of it with other two kinds of four-point diagrams in Figure~\ref{fig:treeA4}, we will obtain two new loop diagrams. This process is shown in Figure~\ref{fig:diagramJacobi}. Note that except for the four-point sub-digram, the rest part of the three diagrams is the same, so their color factors take the form:
\begin{equation}
	C_s=\tilde{f}^{abe}\tilde{f}^{ecd} (\delta\prod\tilde{f}) ,\qquad C_t=\tilde{f}^{bce}\tilde{f}^{eda} (\delta \prod\tilde{f} ),\qquad C_u=\tilde{f}^{ace}\tilde{f}^{ebd} (\delta \prod\tilde{f} ) \,.
\end{equation}
Obviously they satisfy the Jacobi relation:
\begin{equation}\label{eq:loopJacobi}
	C_s=C_t+C_u\,.
\end{equation}
For these diagrams, we can postulate a dual numerator relation as
\begin{equation}\label{eq:dualJocab}
	N_s=N_t+N_u\,.
\end{equation}

As mentioned in the introduction, it is still a conjecture whether such a requirement can be fulfilled at general loop level.
Our strategy is to assume such duality can be achieved and use this structure to construct an ansatz for the loop integrand. 
Then we check the ansatz by physical unitarity-cut constraints. Once the CK-dual ansatz has a solution consistent in all possible unitarity cuts, this then shows that the CK duality can be indeed realized at this order. 
Practically, this conjecture (as long as it works) can also greatly simplify the integrand construction.

We will focus on Sudakov form factor of ${\rm tr}(F^2)$ in pure YM theory, as given in~\eqref{eq:def-sudakovFF}. 
At tree level, the form factor is given as
\begin{equation}
	\label{eq:tree-sudakovFF}
	\hat{\cal F}^{(0)}_{2}(1,2) = C^{a_1,a_2} {\cal F}^{(0)}_{2}(1,2) \,, 
\end{equation}
with color factor 
\begin{equation}
	C^{a_1,a_2}  = {\rm tr}(T^{a_1} T^{a_2}) = \delta^{a_1 a_2}  \,,
\end{equation}
and the color-stripped tree form factor is
\begin{equation}
	{\cal F}^{(0)}_{2}(1,2) = (\varepsilon_1 \cdot \varepsilon_2)(p_1 \cdot p_2)-(\varepsilon_1 \cdot p_2)(\varepsilon_2 \cdot p_1) \,.
\end{equation}
An $l$-loop full-color Sudakov form factor takes the general form as
\begin{equation}\label{eq:generalSudakovFF}
	\hat{\mathcal{F}}_2{}^{(l)}=i^{l} g^{2l}\sum _{\sigma _2} \sum _{\Gamma _i} \int \prod_{j=1}^l {d^dl_j \over (2\pi)^d} \frac{1}{S_i}\frac{C_iN_i}{\
		\Pi_aD_{i,a}}\,,
\end{equation}
with the meaning of each term explained as follows.
The summation over $\sigma_2$ means taking permutation of external legs into account. 
The summation over $\Gamma_i$ means to sum over all possible trivalent graphs. 
$S_i$, coming from overcounting of contributions of the $i$th graph, is the symmetry factor of it. 
$C_i$ is the color factor of the $i$th graph, given as products of structure constants dressed on every vertex. $1/D_{i,a}$ denotes as the $a$th propagator of the $i$th graph. Finally, $N_i$ are the kinematic numerators, which are the main goal of the construction. Reader can find more details of the general strategy in \emph{e.g.}~\cite{Bern:2012uf, Carrasco:2015iwa, Yang:2019vag}.

\subsection{One-loop numerator ansatz}
\begin{figure}[t]
	\centerline{\includegraphics[height=3cm]{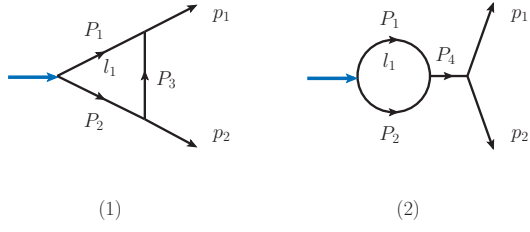} } 
	\caption{One-loop cubic graphs.} 
	\label{fig:1loopConvetnion}
\end{figure}

In this subsection we will construct the ansatz for the one-loop Sudakov form factor.
We start with generating all one-loop cubic graphs. There are two topologies to consider, as shown in Figure~\ref{fig:1loopConvetnion}.

We write the one-loop form factor as
\begin{equation}\label{eq:generalSudakovFF1_loop}
	\hat{\mathcal{F}}_2{}^{(1)}=ig^2\sum _{\sigma _2} \sum _{i=1}^2 \int \frac{d^dl_1}{(2\pi)^d}\frac{1}{S_i}\frac{C_iN_i}{\
		\Pi_aD_{i,a}}\,.
\end{equation}
As in Figure~\ref{fig:1loopConvetnion}, we use $P_i$ to denote the propagator momenta, which will also be used as color indices for each edge. 
We choose the convention of color factors such that the three color indices of each $\tilde{f}$ follow a clockwise direction for each vertex shown in Figure~\ref{fig:1loopConvetnion}. This convention will be used throughout the paper.
Explicitly, the color factors of the graphs in Figure~\ref{fig:1loopConvetnion} are given as
\begin{equation}
	\label{eq:1loopcolor}
	C_{1}=\delta^{P_1P_2}\tilde{f}^{P_1p_1P_3}\tilde{f}^{P_3 p_2P_2} \,,\qquad C_{2}=\delta^{P_1P_2}\tilde{f}^{P_2 P_1P_4}\tilde{f}^{P_4p_1p_2}\,.
\end{equation}
We would like to point out that  $C_{2}$ is zero since the product of $\delta^{P_1P_2}$ and $\tilde{f}^{P_2 P_1P_4}$ vanishes. However, as we will discuss below, it is important to keep it in the form of~\eqref{eq:1loopcolor} for getting the correct sign factors for the dual Jacobi relations.
The two kinematic numerators are defined as  
\begin{equation}\label{eq:1loopnumconvention}
	N_1[l_1, p_1,p_2] \,, \qquad  N_2[l_1,p_1,p_2]\,,
\end{equation}
which are the main functions to be solved.

\subsubsection{Generating CK relations}\label{chap:1-loopCKrelation}
In this part we use the one-loop form factor to describe in detail how to generate dual Jacobi relations that will provide linear relations for the numerators.

\begin{figure}[t]
	\centerline{\includegraphics[height=5.4cm]{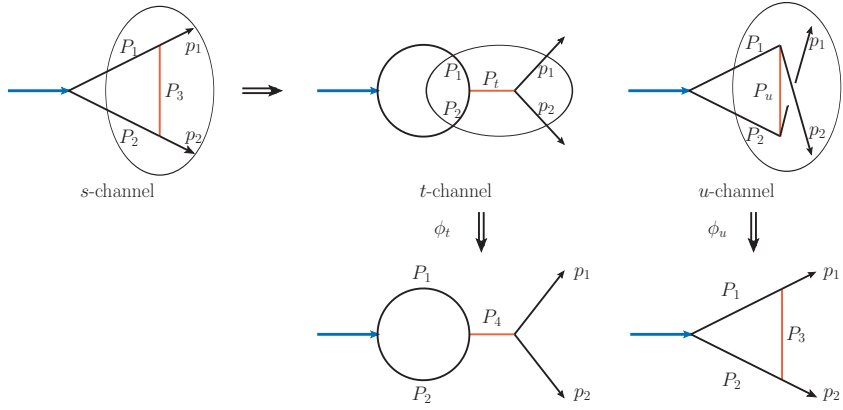} } 
	\caption{Map from CK operation results to convention.} 
	\label{fig:mapToColorConvention}
\end{figure}

We start with picking up a propagator in the trivalent graph which will play the role of $s$-channel propagator in Figure~\ref{fig:diagramJacobi}.
In order to apply the dual Jacobi relation, such a propagator should not be connected to the operator leg because the operator vertex is not dressed with a structure constant $\tilde{f}^{abc}$.
For the first topology in Figure~\ref{fig:1loopConvetnion}, the only propagator which can be considered is denoted by red color in Figure~\ref{fig:mapToColorConvention}. 
The Jacobi relation leads to the $t$-channel and $u$-channel graphs in  Figure~\ref{fig:mapToColorConvention}.
At one loop this will be the only dual Jacobi relation one needs to consider.

\paragraph{Color Jacobi relation.}
We first consider the color Jacobi relation which will be important to determine the sign factors in the dual Jacobi relation for the kinematic numerators.
We set $C_{s}$ as $C_1$ in~\eqref{eq:1loopcolor} and 
define the other two color factors as
\begin{equation}\label{eq:1loopCKoperationColor}
	C_s=\delta^{P_1P_2}\tilde{f}^{P_1p_1P_3}\tilde{f}^{P_3 p_2P_2},\quad C_t=\delta^{P_1P_2}\tilde{f}^{p_1p_2P_t}\tilde{f}^{P_tP_2P_1},\quad C_u=\delta^{P_1P_2}\tilde{f}^{P_1p_2P_u}\tilde{f}^{P_up_1P_2},
\end{equation}
which satisfy 
\begin{equation}
	C_s=C_t+C_u \,. 
\end{equation}
To extract the correct sign factors for the dual Jacobi relation, we can map the color indices in $C_{t/u}$ to our convention in the form of~\eqref{eq:1loopcolor}.
As shown in Figure~\ref{fig:mapToColorConvention}, the $t$- and $u$-channel topologies are isomorphic to the bubble and triangle topologies respectively. 
We define the map of color indices as $\phi_{t/u}$ shown in Figure~\ref{fig:mapToColorConvention} 
which are
\begin{equation}\label{eq:oneloopMapOperator}
		\phi_t=\left\{P_t\rightarrow P_4\right\}\,,\qquad
		\phi_u=\left\{P_1\leftrightarrow P_2,P_u\rightarrow P_3\right\}\,.
\end{equation}
As a result we will obtain 
\begin{equation}\label{eq:oneloopJocabi}
	\begin{split}
		&C_t=\phi_t\left[C_t\right]=\delta^{P_1P_2}\tilde{f}^{p_1p_2P_4}\tilde{f}^{P_4P_2P_1}=\left(+\right)C_{2}\,,\\
		&C_u=\phi_u\left[C_u\right]=\delta^{P_2P_1}\tilde{f}^{P_2p_2P_3}\tilde{f}^{P_3p_1P_1}=\left(+\right)C_{1}\,.
	\end{split}
\end{equation}
Note that the equality after $C_{t/u}$ holds because $\phi_{t/u}$ only rename color indices. Then we can directly compare mapped $C_{t/u}$ with $C_{i}$. The two plus signs are the results of the comparison.
We have 
\begin{equation}\label{eq:color1loopjacobi}
	C_{1}=C_s = C_t+C_u=\left(+\right)C_{2}+\left(+\right)C_{1} \,.
\end{equation}
The same signs will be used for the kinematic numerators. We emphasize again that although $C_t=C_2=0$ is correct in value, we need to use its form as in \eqref{eq:oneloopJocabi} to get the wanted sign. 
See also \cite{Lin:2021qol} for further examples and discussion on this point.

\paragraph{Dual Jacobi relation.} 
Next we derive the dual Jacobi relation for the numerator functions~\eqref{eq:1loopnumconvention}. 
Mapping to the momenta in the $t$- and $u$-channel diagrams in Figure~\ref{fig:mapToColorConvention}, 
we have the dual Jacobi relation:
\begin{equation}
	\begin{split}
		N_1[l_1,p_1,p_2]&=(+)N_2[P_1,p_1,p_2]+(+)N_1[P_2,p_1,p_2]\\
		&= N_2[l_1,p_1,p_2]+N_1[p_1+p_2-l_1,p_1,p_2] \,,
	\end{split}
\end{equation}
where the two plus signs in the first equation are in accord with the signs in the color relation~\eqref{eq:color1loopjacobi}.

\paragraph{Choose master numerator.} 

We can represent $N_2$ by a linear combination of $N_1$ as
\begin{equation}\label{eq:expressN2byN1}
	N_2\left[l_1,p_1,p_2\right]=N_1\left[l_1,p_1,p_2\right]-N_1\left[p_1+p_2-l_1,p_1,p_2\right]\,,
\end{equation}
We will choose $N_1$ as the master numerator. Any groups of numerators which can express all numerators through CK relations can be called a set of \emph{CK master numerators}. Corresponding cubic graphs of these numerators are called \emph{CK master topology}. 
In this one-loop case, there is a unique choice of master numerator as $N_1$. However, in general the choice of master numerators are not unique, such as in the two-loop case discussed below. 
Usually we tend to choose simple topologies (such as planar topologies and symmetric topologies) and also minimize the number of master topologies, so that the ansatz construction will be simplified.

\subsubsection{Constructing ansatz for the master numerator}

\begin{figure}[t]
	\centerline{\includegraphics[height=2.5cm]{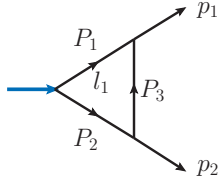} }
	\caption{One-loop master graph and the labeling of its propagators.} 
	\label{fig:1looppropagator}
\end{figure}

We need to further construct an ansatz for the master numerator $N_1$.
The numerator is given as a polynomial in terms of following Lorentz product variables:
\begin{equation}
	\begin{split}
		& D_i=P_i^2,\qquad  s=(p_1+p_2)^2,\qquad \mathcal{E}=\varepsilon_1\cdot\varepsilon_2,\\
		& \mathcal{K}_{ij}=\varepsilon_i\cdot k_j,\quad \text{with}\ k_1=p_1,k_2=p_2,k_3=P_1=l_1\,,
	\end{split}
\end{equation}
where the momenta $P_i$ for propagators are shown in Figure~\ref{fig:1looppropagator}.
The complete basis is
\begin{equation}\label{eq:oneLoopBasis}
	\{ D_1,D_2,D_3,s,\mathcal{E}, \mathcal{K}_{12}, \mathcal{K}_{13}, \mathcal{K}_{21}, \mathcal{K}_{23} \} .
\end{equation}

We impose the power counting condition following Feynman diagrams. The one-loop diagram contains two trivalent-vertices and an operator vertex. Each trivalent-vertex contributes mass dimension one to the numerator, and the operator vertex contributes mass dimension two. For the complete Feynman rule of the form factor, see \emph{e.g.}~\cite{Koukoutsakis:2003nba}. So numerator needs to be mass dimension four in total.  
In addition, the numerator must depend on each polarization vector linearly. After collecting all permitted monomials, the ansatz of $N_1$ can be written as
\begin{align}\label{eq:1loopN1ansatz}
	N_{1}\left[l_1,p_1,p_2\right]=&a_1 D_1^2 \mathcal{E}+a_{2} D_2^2 \mathcal{E}+a_3 D_3^2 \mathcal{E}+a_4 D_1 D_2 \mathcal{E}+a_5 D_1 D_3 \mathcal{E}+a_6 D_2 D_3 \mathcal{E} \\
	&+a_{7} D_1 \mathcal{K}_{13} \mathcal{K}_{23}+a_{8} D_1 \mathcal{K}_{12} \mathcal{K}_{23}+a_{9} D_1 \mathcal{K}_{13} \mathcal{K}_{21}+a_{10} D_1\mathcal{K}_{12}\mathcal{K}_{21}+a_{11} D_1 s \mathcal{E} \notag\\
	&+a_{12} D_2 \mathcal{K}_{13} \mathcal{K}_{23}+a_{13} D_2 \mathcal{K}_{12} \mathcal{K}_{23}+a_{14} D_2 \mathcal{K}_{13} \mathcal{K}_{21}+a_{15} D_2 \mathcal{K}_{12} \mathcal{K}_{21}+a_{16} D_2 s \mathcal{E} \notag\\
	&+a_{17} D_3 \mathcal{K}_{13} \mathcal{K}_{23}+a_{18} D_3 \mathcal{K}_{12} \mathcal{K}_{23}+a_{19} D_3 \mathcal{K}_{13} \mathcal{K}_{21}+a_{20} D_3 \mathcal{K}_{12} \mathcal{K}_{21}+a_{21} D_3 s \mathcal{E} \notag\\
	&+a_{22} \mathcal{K}_{13} \mathcal{K}_{23} s+a_{23} \mathcal{K}_{12} \mathcal{K}_{23} s+a_{24} \mathcal{K}_{13} \mathcal{K}_{21} s+a_{25} \mathcal{K}_{12} \mathcal{K}_{21} s +a_{26} s^2 \mathcal{E}\,.\notag
\end{align}
There are 26 monomials in total. Notice that for completeness, we include terms containing $D_{1,2}$, such as $a_{7} D_1\mathcal{K}_{13} \mathcal{K}_{23}$ and $a_4 D_1 D_2 \mathcal{E}$, which will reduce to massless bubbles or tadpole, and they are zero after integration. As we will see, 
the coefficients of these kinds of terms can not be determined by the unitarity cut,
and one can also exclude such terms in the ansatz from the beginning. 
However, this is not possible in the two-loop case, which will be discussed in Section~\ref{sec:2loopunitarity}.

\begin{figure}[t]
	\centerline{\includegraphics[height=2.5cm]{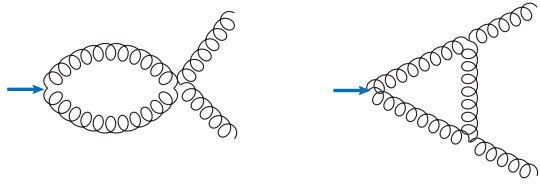} }
	\caption{Feynman diagrams which have $d$ dependence in the numerators.} 
	\label{fig:1loopdexpansion}
\end{figure}

There is another important comment on the coefficients $a_i$. From the Feynman diagram calculation, two Feynman graphs as shown in Figure~\ref{fig:1loopdexpansion} can produce terms that are linear in the space-time dimension parameter $d$ in the numerator, due to the contraction of metric $\eta^{\mu\nu}\eta_{\mu\nu}=\delta^{\mu}_\mu = d$.
Therefore, we expect that the coefficients will in general be degree-one polynomials of $d$ as
\begin{equation}\label{eq:oneloopAnsatzExpansion}
	a_j=a_{j,0}+d\, a_{j,1}\,,
\end{equation}
where $a_{j,0}$ and $a_{j,1}$ are pure rational numbers.

\subsubsection{Symmetrization of the ansatz}
\label{sec:sym1loop}
Here we would like to impose a further constraint on the ansatz such that each numerator should respect the symmetry property of the corresponding diagram.
More concretely, symmetry conditions come from graph self-isomorphism and we require that $C_{i} N_i$ goes back to itself after the symmetry transformations.

Considering the first graph in Figure~\ref{fig:1loopConvetnion}, along the horizontal axis, there is a symmetry transformation:
\begin{equation}
	\textbf{s}_1=\left\{P_1\leftrightarrow P_2, \ p_1\leftrightarrow p_2\right\}\,.
\end{equation}
For the second graph in Figure~\ref{fig:1loopConvetnion}, it has two symmetries, by exchanging bubble propagators or exchanging external legs, and they correspond to the following transformations respectively:
\begin{equation}\
		\textbf{s}_2=\left\{P_1\leftrightarrow P_2\right\}\,, \qquad 
		\textbf{s}_3=\left\{p_1\leftrightarrow p_2\right\}\,.
\end{equation}
These symmetry transformations impose the following three constraints:
\begin{equation}\label{eq:1loopsym}
	\begin{split}
		C_{1}N_1\left[P_1,p_1,p_2\right]=\textbf{s}_1\left[C_{1}N_1\left[P_1,p_1,p_2\right]\right]
		&=\delta^{P_2P_1}\tilde{f}^{P_2p_2P_3}\tilde{f}^{P_3 p_1P_1}N_1\left[P_2,p_2,p_1\right]\\
		&=C_{1}N_1\left[P_2,p_2,p_1\right]\,,\\
		C_{2}N_2\left[P_1,p_1,p_2\right]=\textbf{s}_2\left[C_{2}N_2\left[P_1,p_1,p_2\right]\right]
		&=\delta^{P_2P_1}\tilde{f}^{P_1P_2P_4}\tilde{f}^{P_4p_1p_2}N_2\left[P_2,p_1,p_2\right]\\
		&=-C_{2}N_2\left[P_2,p_1,p_2\right]\,,\\
		C_{2}N_2\left[P_1,p_1,p_2\right]=\textbf{s}_3\left[C_{2}N_2\left[P_1,p_1,p_2\right]\right]
		&=\delta^{P_1P_2}\tilde{f}^{P_2P_1P_4}\tilde{f}^{P_4p_2p_1}N_2\left[P_1,p_2,p_1\right]\\
		&=-C_{2}N_2\left[P_1,p_2,p_1\right]\,.
	\end{split}
\end{equation}
As discussed below~\eqref{eq:oneloopJocabi}, we can not simply set $C_2$ to 0 but should use the form with color indices shown explicitly.
In terms of the numerator functions, they are
\begin{equation}
	\begin{split}
		N_1\left[l_1,p_1,p_2\right]&=N_1\left[p_1+p_2-l_1,p_2,p_1\right]\,,\\
		N_2\left[l_1,p_1,p_2\right]&=-N_2\left[p_1+p_2-l_1,p_1,p_2\right]\,,\\
		N_2\left[l_1,p_1,p_2\right]&=-N_2\left[l_1,p_2,p_1\right]\,.
	\end{split}
\end{equation}
These conditions will solve for 9 coefficients in $N_{1}$, leaving 17 coefficients unsolved.

\subsection{Two-loop numerator ansatz}\label{sec:2loopansatz}
In this subsection we go on to construct CK-dual ansatz for the two-loop Sudakov form factor. 
Since the procedure is similar to the one-loop case, we will be brief in the discussion and mainly focus on the new features at two loops.

\begin{figure}[t]
	\centerline{\includegraphics[height=7.cm]{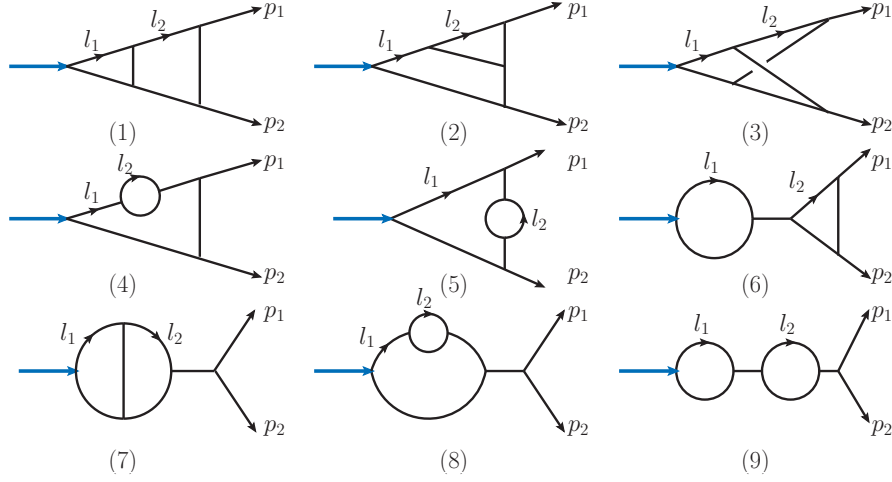} } 
	\caption{Two-loop cubic graphs without massless bubble and tadpole.} 
	\label{fig:2looptopo}
\end{figure}

\begin{figure}[t]
	\centerline{\includegraphics[height=5.4cm]{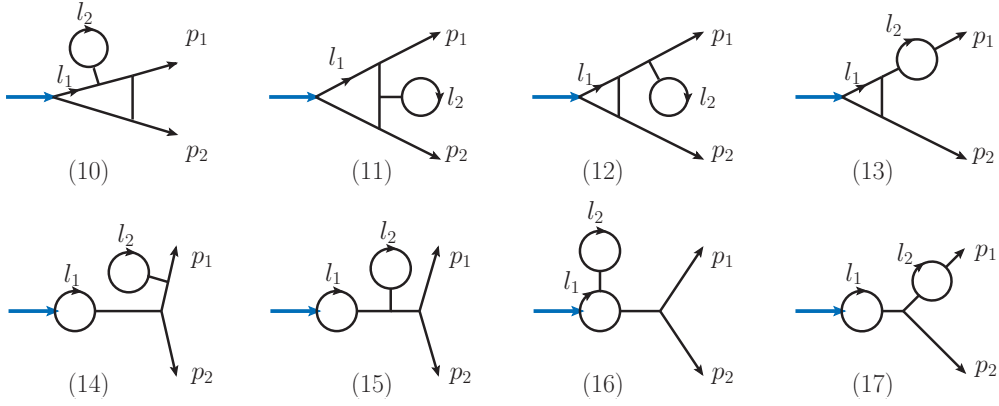} } 
	\caption{Two-loop cubic graphs with massless bubble and tadpole.} 
	\label{fig:2looptadpoleandmassless}
\end{figure}

To begin with, we generate all possible cubic graphs. There are 17 different graphs which are shown in Figure~\ref{fig:2looptopo} and Figure~\ref{fig:2looptadpoleandmassless}.
In particular, eight of them contain massless bubble or tadpole sub-graphs, which are all collected in Figure~\ref{fig:2looptadpoleandmassless}. 
The two-loop form factor can be given the following ansatz form:
\begin{equation}\label{eq:generalSudakovFF2_loop}
	\hat{\mathcal{F}}_2{}^{(2)}=i^2g^4\sum _{\sigma _2} \sum _{i=1}^{17} \int \prod_{j=1}^2 \frac{d^dl_j}{(2\pi)^d}\frac{1}{S_i}\frac{C_iN_i}{\
		\Pi_aD_{i,a}}\,.
\end{equation}     
As mentioned, the convention of color factors are chosen such that the color indices of each $\tilde{f}^{abc}$ follow clockwise direction for the corresponding vertex in Figure~\ref{fig:2looptopo} and Figure~\ref{fig:2looptadpoleandmassless}. The kinematic numerators are defined as functions $N_i[l_1, l_2, p_1, p_2]$, where $l_{1,2}$ are labeled in figures.

\subsubsection{Generating CK relations}
Following the strategy in Section~\ref{chap:1-loopCKrelation}, we have generated all dual Jacobi relations in Appendix~\ref{app:CKrelation}. 
As mentioned before, the choice of the master topologies is not unique.
We will choose the numerators of the first two graphs in Figure~\ref{fig:2looptopo} as master numerators, which are both planar. 

\begin{figure}[t]
	\centerline{\includegraphics[height=2.6cm]{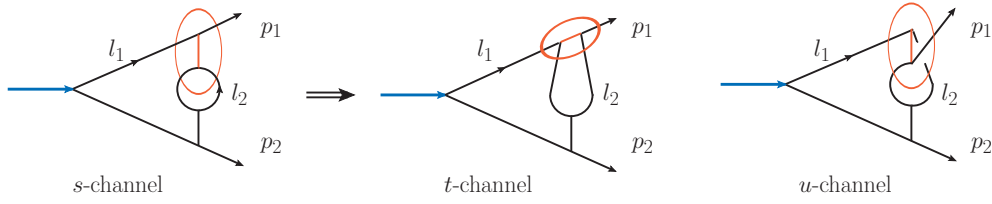} } 
	\caption{CK operation corresponding to the the third CK relation.} 
	\label{fig:2loopCKeg}
\end{figure}

Using a set of 15 dual Jacobi relations, one can obtain all other 15 numerators from the two master numerators. An explicit set of such relations can be give as 
\begin{align}\label{eq:twoloopCK}
	N_3&=N_1-N_2\,,\\
	N_4&=N_2+N_2 \left[l_1,l_1-l_2,p_1,p_2\right]\,,\\
	N_5&=N_2\left[l_1,p_1-l_2,p_1,p_2\right]+N_2\left[l_1,l_1+l_2,p_1,p_2\right]\,, \label{eq:dualJacobiexample}\\
	N_6&=N_1-N_1 \left[p_1+p_2-l_1,l_2,p_1,p_2\right]\,, \label{eq:N6}\\
	N_7&=N_1-N_1 \left[l_1,l_2,p_2,p_1\right]\,, \label{eq:N7}\\
	N_8&=N_7+N_7 \left[l_1,l_1-l_2,p_1,p_2\right]\,,\\
	N_9&=N_7+N_7 \left[l_1,p_1+p_2-l_2,p_1,p_2\right]\,, \label{eq:N9}\\
	N_{10}&=N_4-N_4\left[l_1,-l_2,p_1,p_2\right]\,,\\
	N_{11}&=-N_5+N_5\left[l_1,-l_2,p_1,p_2\right]\,,\\
	N_{12}&=-N_{13}+N_{13}\left[l_1,-l_2,p_1,p_2\right]\,,\\
	N_{13}&=N_2+N_2\left[l_1,p_1-l_2,p_1,p_2\right]\,,\\
	N_{14}&=N_{17}-N_{17}\left[l_1,-l_2,p_1,p_2\right]\,,\\
	N_{15}&=N_9-N_9\left[l_1,-l_2,p_1,p_2\right]\,,\\
	N_{16}&=N_8-N_8\left[l_1,-l_2,p_1,p_2\right]\,,\\
	N_{17}&=N_6+N_6\left[l_1,p_1-l_2,p_1,p_2\right]\,,
\end{align}
where $N_i$ represent $N_i[l_1,l_2,p_1,p_2]$ for brevity.
The first seven relations can generate other seven numerators in Figure~\ref{fig:2looptopo}, while the other eight relations can be used to generate all numerators in  Figure~\ref{fig:2looptadpoleandmassless}. As an example, we demonstrate the relation \eqref{eq:dualJacobiexample} as diagrams in Figure~\ref{fig:2loopCKeg}.

\subsubsection{Constructing ansatz for master numerators}
\label{eq:2loopNumansatz}
\begin{figure}[t]
	\centerline{\includegraphics[height=2.4cm]{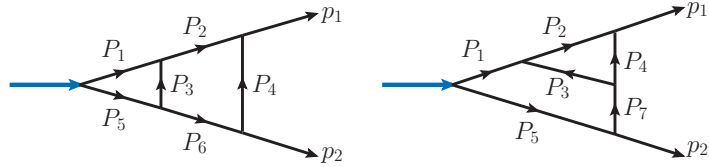} }
	\caption{Two-loop propagator basis.} 
	\label{fig:2looppropagator}
\end{figure}

Next we construct the ansatz for master numerators $N_1$ and $N_2$. The momenta for propagator basis are labeled as $P_i$ in Figure~\ref{fig:2looppropagator}.

Define the two-loop Lorentz products as
\begin{equation}
	\begin{split}
		& D_i=P_i^2\,, \qquad s=(p_1+p_2)^2\,, \qquad \mathcal{E}=\varepsilon_1\cdot\varepsilon_2\,, \\
		& \mathcal{K}_{ij}=\varepsilon_i\cdot  k_j, \quad \text{with}\ k_1=p_1,k_2=p_2,k_3=P_1=l_1,k_4=P_2=l_2\,.
	\end{split}
\end{equation}
The numerators are given as polynomials in terms of following Lorentz product basis, which contains 15 elements:
\begin{equation}\label{eq:twoLoopBasis}
	\{  D_1,D_2,D_3,D_4,D_5,D_6,D_7,s,\mathcal{E}, \mathcal{K}_{12}, \mathcal{K}_{13}, \mathcal{K}_{14}, \mathcal{K}_{21}, \mathcal{K}_{23}, \mathcal{K}_{24}\}.
\end{equation}

We use power counting of the Feynman diagram: there are 4 three-vertices and an operator vertex, we require the numerators to have mass dimension 6. 
The ansatz for each master numerator turns out to have 444 different monomials in the most general form:
\begin{equation}
	\label{eq:2loopansatzN1N2}
	N_1 = \sum_{i=1}^{444} a_i M_i \,, \qquad N_2 = \sum_{i=1}^{444} b_i M_i \,,
\end{equation}
where $M_i$ are the monomials similar in the one-loop ansatz \eqref{eq:1loopN1ansatz}.

\begin{figure}[t]
	\centerline{\includegraphics[height=2.5cm]{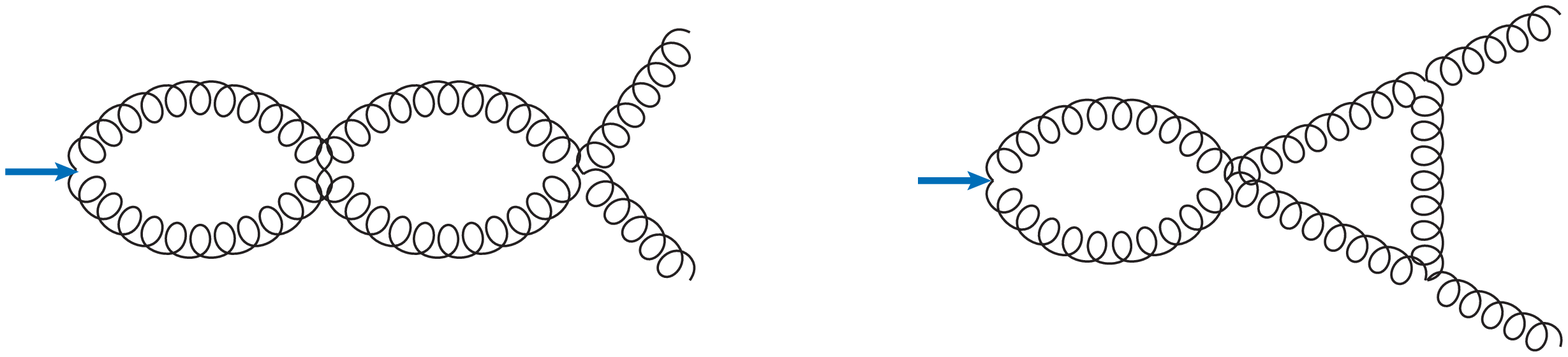} } 
	\caption{Feynman diagrams which have contribution of $d^2$.} 
	\label{fig:d^2_diag}
\end{figure}

As in the one-loop case, the numerators from Feynman diagrams can also depend on $d$. An inspection of Feynman diagrams shows that only two Feynman diagrams, as shown in Figure~\ref{fig:d^2_diag}, can produce terms proportional to $d^2$, coming from metric contractions $\eta^{\mu\nu}\eta_{\mu\nu}\eta^{\mu'\nu'}\eta_{\mu'\nu'}= d^2$. 
Thus we expect the coefficients to be in general degree-two polynomials in $d$ as
\begin{equation}\label{eq:twoloopCoefficientExpansion}
		a_{k} =a_{k,0}+d\, a_{k,1}+d^2\, a_{k,2}\,,\qquad
		b_{k} =b_{k,0}+d\, b_{k,1}+d^2\, b_{k,2}\,,
\end{equation}
where $a_{k,\alpha}$ and $b_{k,\alpha}$ are pure rational numbers.

\paragraph{$d^2$-dependence.} One can note that the two Feynman diagrams in Figure~\ref{fig:d^2_diag} have special loop structures that they are related to trivalent topologies (1), (6), (7) and (9) in Figure~\ref{fig:2looptopo}.
Thus one may expect that only the numerators of these four topologies 
$\{N_1, N_6, N_7, N_9\}$ have $d^2$ dependence.
Using dual Jacobi relations, one can find that there is a chain of relations \eqref{eq:N6},  \eqref{eq:N7}, and  \eqref{eq:N9}, such that the numerators $N_6, N_7, N_9$ can be generated by only using $N_1$ (which is one of the master numerators). 
This raises the question that if it is possible only to let $N_1$ contain $d^2$ terms and set $d^2$ terms of the second master numerator $N_2$ equal zero. If possible, this may be used to simplify the ansatz and reduce the number of free parameters.
However, as we will see in Section~\ref{sec:2loopunitarity}, this is in general not possible. The unitarity-cut constraints together with CK duality require us to include $d^2$ terms in $N_2$.

\paragraph{Reduced tadpoles and massless bubbles.} 
In the ansatz consisting of full 444 terms in \eqref{eq:2loopansatzN1N2}, we include monomials that can reduce the maximal topologies to sub-topologies that contain tadpoles and massless bubbles. 
Such terms correspond to scaleless integrals and will not contribute to the final form factor result. Moreover they are not detectable by unitarity cuts. 
Thus we may wonder if one can exclude these terms from the ansatz for the master numerators. 
If possible, this can also simplify the ansatz for master numerators from the beginning. 
However, as we will see in Section~\ref{sec:2loopunitarity}, this is in general not possible. 
The unitarity-cut constraints together with CK duality require us to include such terms in the master numerators.

\subsubsection{Symmetrization of the ansatz}

\label{sec:sym2loop}
\begin{figure}[t]
	\centerline{\includegraphics[height=2.8cm]{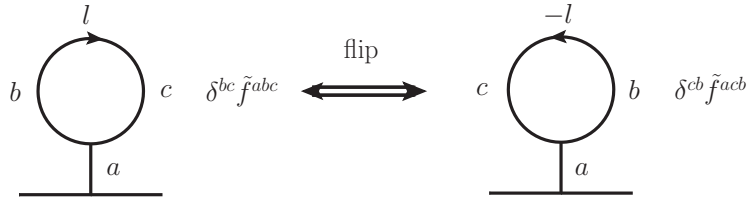} } 
	\caption{Symmetry of tadpole.} 
	\label{fig:tadpole_sym}
\end{figure}

Finally, like the one-loop case discussed in Section~\ref{sec:sym1loop}, we impose symmetry constraints on the numerators. 
Since the tadpole topologies appear in the two-loop case for the first time, we discuss their feature in more detail. Each tadpole subgraph has the flip symmetry as shown in Figure~\ref{fig:tadpole_sym}. Under the flip symmetry, the momentum flowing in the tadpole will change its direction and the color factor will produce an additional minus sign. Thus, using the topology (10) in Figure~\ref{fig:2looptadpoleandmassless} as an example,  one has the constraint on the numerator
\begin{equation}
	N_{10}\left[l_1,-l_2,p_1,p_2\right]=-N_{10}\left[l_1,l_2,p_1,p_2\right]\,,
\end{equation} 
from which $N_{10}$ must be made up only by terms with odd powers of tadpole-loop momentum.
We point out that this symmetry property is actually automatically satisfied if the tadpole numerators are generated through dual Jacobi relations, see Section~\ref{sec:constraintstadpole} for further discussion. 

It turns out to be that 508 parameters of $N_1$ and $N_2$ can be solved after using symmetry constraints, and there are 380 parameters left. We point out that after this step, the numerators also satisfy all dual Jacobi relations as listed in Appendix~\ref{app:CKrelation}.

\section{Solving CK ansatz}\label{chap:unitarity}
\label{sec:unitaritycuts}
Having obtained the CK ansatz, now we apply the unitarity cuts to constrain the ansatz so that it provides the physical result.
We stress that one needs to apply $d$-dimensional cuts to get the complete result, since we are studying the pure YM theory. The feature that the ansatz numerators depend on the dimension parameter $d$ also implies that $d$-dimensional cuts are necessary.

\subsection{Review of $d$-dimensional cut}
The central idea of the unitarity method \cite{Bern:1994zx,Bern:1994cg,Britto:2004nc} is that by setting internal propagators to be on-shell as
\begin{equation}\label{eq:oneloopNuExpansion}
	\frac{i}{l^2} \stackrel{\rm cut}{\longrightarrow} 2\pi\delta_{+}(l^2) \,,
\end{equation}
the amplitude or form factor will be factorized as product of lower-order amplitudes or form factors, such as
\begin{equation}
	{\cal F}^{(l)} |_{\rm cuts} = \sum_{\rm helicities} \prod \textrm{(tree-level blocks)} \,.
\end{equation}
The physical result must be consistent in all possible cut channels. 
Note that since there is no sub-leading color contribution for Sudakov form factor at one and two loops, it is enough to consider planar cuts.
(See also \cite{Lin:2021qol} for discussion on the non-planar cuts.)

To apply $d$-dimensional planar cuts, the building blocks are color-stripped tree amplitudes or form factors which can be calculated by Feynman diagrams. All the expressions appear as Lorentz products which are valid in $d$ dimensions. When we multiply them together, we need to sum over all the helicities, and in $d$-dimensional cut this operation corresponds to the contraction of the polarization vectors for the internal gluons by using the following rule:

\begin{equation}\label{eq:contraction_rule}
	\sum _{\text{helicities}} \; \varepsilon^{\mu}(l)\varepsilon^{\nu}(l) =\eta^{\mu\nu}\;-\ \frac{l^{\mu}\xi^{\nu}+l^{\nu}\xi^{\mu}}{l\cdot \xi},
\end{equation}
where
the $\xi^{\mu}$ is a light-like reference momenta and the result after summation should be independent of the choice of it. 
By matching the resulting tree product expression and the cut of the ansatz integrand, one can fix the coefficients in the ansatz.

We checked that after the helicity sum, the $\xi$-dependent terms in the expression vanish. As another check, when we set $d=4$, the tree products match the result obtained by spinor helicity formalism. 
Compared to the 4-dimensional cut using spinor helicity formalism, $d$-dimensional cuts can capture all terms valid for general dimensions.

\subsection{One-loop solution}
\label{sec:1loopsolution}

\begin{figure}[t]
	\centerline{\includegraphics[height=2.7cm]{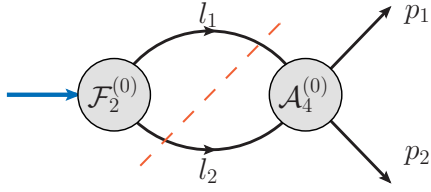}}
	\caption{Double cut for one-loop two-point form factor.} 
	\label{fig:Double cut for one loop two point form factor}
\end{figure}

For the one-loop case, there is only one cut we need to consider, which is the double cut in Figure~\ref{fig:Double cut for one loop two point form factor}. The one-loop Sudakov form factor will be factorized as the products of the two-point tree form factor and the four-point tree amplitude:
\begin{equation}
	\label{eq:oneloopCut}
	\mathcal{F}^{(1)}_2|_{s_{12}\textrm{-cut}}\text{=$\int $d}\text{PS}_2\text{  }\sum _{\text{helicities}}  \mathcal{F}_2^{(0)}\left(l_1,l_2\right)\mathcal{A}_4^{(0)}{(1,2,-}l_2{,-}l_1) \,,
\end{equation}
where
the integral measure is the two-particle phace space measure which is defined as
\begin{equation}
	d\text{PS}_2=\frac{d^Dl_1}{2\pi ^D}2\pi \delta _+\left(l_1^2\text{)2}\pi \delta _+\text{((}p_1+p_2-l_1)^2\right) .
\end{equation}

As discussed in the previous section, after the symmetry constraints 
there are 17 monomial coefficients left in the ansatz. When we match the CK ansatz under the cut and the tree product as in \eqref{eq:oneloopCut}, we can further fix 7 coefficients and 10 of them remain unsolved.
These coefficients can not be fixed by symmetry and unitarity. 
We find that in this solution space, we can let the numerator of the second topology be zero, whose color factor is zero and will not contribute to the final result. When we apply this extra constraint, there will be only 6 free coefficients.

We present the final numerator of the triangle topology by separating it into two parts as
\begin{equation}
	N_{1}[l_1,p_1,p_2] = N_{1,0}[l_1,p_1,p_2] + N_{1,1}[l_1,p_1,p_2] \,,
\end{equation} 
where $N_{1,0}$ contains no free parameters and all the monomials with six unfixed coefficients are collected in $N_{1,1}$. They are explicitly given as
\begin{align}
	N_{1,0} =&
	-s^2 \mathcal{E}+2 \mathcal{K}_{12} \mathcal{K}_{21} s+\epsilon (2 D_3^2 \mathcal{E}-2 D_1 D_3 \mathcal{E}-2 D_2 D_3 \mathcal{E}-4 D_1 \mathcal{K}_{13} \mathcal{K}_{21}-4 D_2 \mathcal{K}_{12} \mathcal{K}_{21} \nonumber\\
	&+4 D_2 \mathcal{K}_{12} \mathcal{K}_{23}+4 D_3 \mathcal{K}_{13} \mathcal{K}_{21}-4 D_3 \mathcal{K}_{12} \mathcal{K}_{23}+2 D_3 s \mathcal{E}+4 \mathcal{K}_{13} \mathcal{K}_{21} s-4 \mathcal{K}_{13} \mathcal{K}_{23} s) , 	\label{eq:N10} \\
	N_{1,1}=&
	\, a_1 \left(D_1^2+D_2^2\right) \mathcal{E}+a_4 D_1 D_2 \mathcal{E}+a_{10} \left(D_1+D_2\right) \mathcal{K}_{12} \mathcal{K}_{21}+a_{11} \left(D_1+D_2\right) s \mathcal{E} \nonumber\\
	&+a_{7} \left(D_1 \mathcal{K}_{13} \mathcal{K}_{23}+D_2 \mathcal{K}_{12} \mathcal{K}_{21}+D_2 \mathcal{K}_{13} \mathcal{K}_{23}-D_2 \mathcal{K}_{13} \mathcal{K}_{21}-D_2 \mathcal{K}_{12} \mathcal{K}_{23}\right)\nonumber\\
	&+a_{8} \left(D_1 \mathcal{K}_{13} \mathcal{K}_{21}+D_1 \mathcal{K}_{12} \mathcal{K}_{23}+2 D_2 \mathcal{K}_{12} \mathcal{K}_{21}-D_2 \mathcal{K}_{13} \mathcal{K}_{21}-D_2 \mathcal{K}_{12} \mathcal{K}_{23}\right)\,.\label{eq:N11}
\end{align}
In \eqref{eq:N10}, except the first two terms, all other terms are proportional to $\epsilon=(4-d)/2$ and they can not be fixed in the 4-dimensional cut. For the terms depending on free coefficients collected in \eqref{eq:N11}, we can easily see that all of them are proportional to $D_1$ or $D_2$, which will reduce to topologies with massless bubble or tadpole. They can not be detected by unitarity cuts and thus they remain as free parameters. Since they are scaleless integrals, they vanish after integration.

\subsection{Two-loop solution}
In this subsection, we will constrain the two-loop CK ansatz. We will first apply unitarity cuts. As we will see, there is a large solution space. Next we will discuss possible further constraints to reduce the solution space from tadpoles and massless bubbles.
Finally, we discuss some features of the solution space.

\subsubsection{Unitarity constraints}
\label{sec:2loopunitarity}

\begin{figure}[t]
	\centerline{\includegraphics[height=4.5cm]{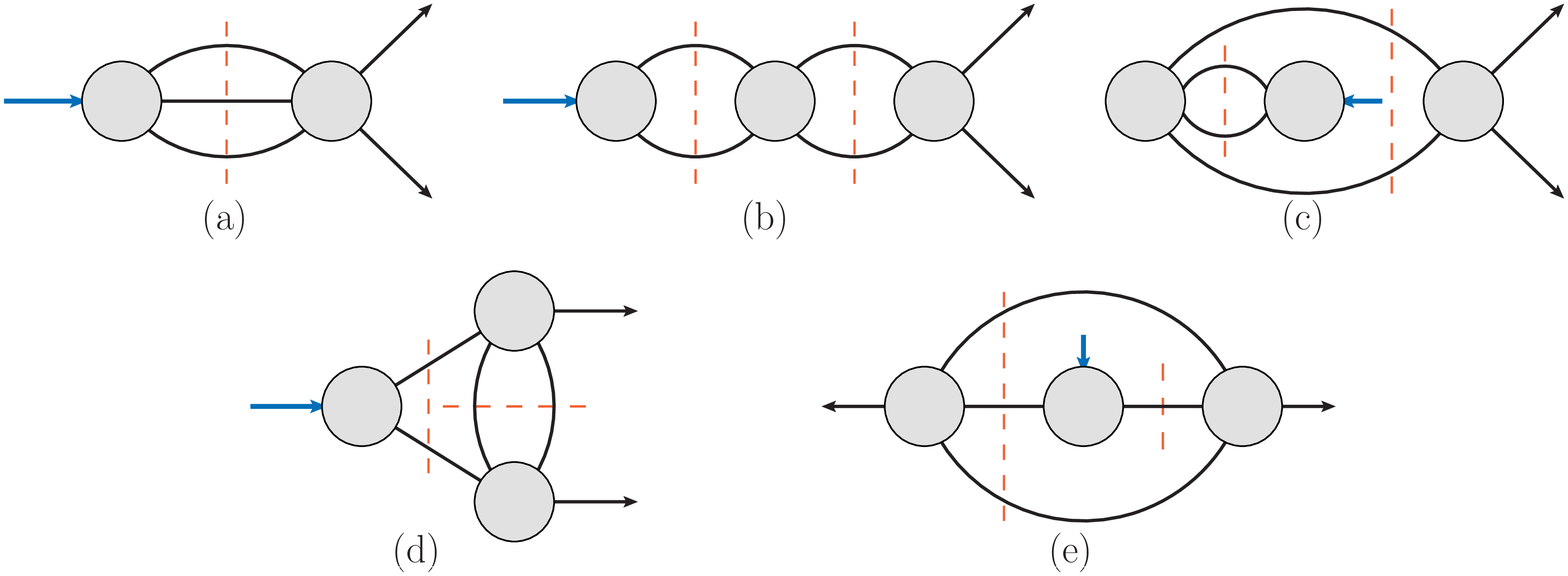}} 
	\caption{A spanning set of cuts for two-loop Sudakov form factor.} 
	\label{fig:2loop_spanning_set_cuts}
\end{figure}

At two loops, it is necessary to consider several different cuts. A complete spanning set of planar cuts are shown in Figure~\ref{fig:2loop_spanning_set_cuts}.
We mention that cut (b) and cut (c) give the same constraint because of the cyclic and reflection symmetry of the 4-point tree amplitude.

As discussed in Section~\ref{sec:2loopansatz}, after requiring all the numerators to have the same symmetry as their topologies, we are able to constrain a lot of parameters and there will be 380 coefficients left.
After applying all the cuts, there are 235 coefficients that remain unsolved. 
All of the remaining parameters should vanish after the integral reduction, which we have checked to be so.
Before considering further constraints, 
let us address the two questions we raised during the construction of ansatz in Section~\ref{eq:2loopNumansatz}.

\paragraph{On $d^2$-dependence.}  First we consider the question if one can set the $d^2$ terms in $N_2$ to be zero. 
In the unitarity calculation we find that only in the cut of Figure~\ref{fig:2loop_spanning_set_cuts}(b), the tree product will provide non-zero $d^2$ terms, which is consistent with the Feynman diagram analysis. Topologies that contribute to cut-(b) are (1), (6), (7) and (9) in Figure~\ref{fig:2looptopo}. 
However, 
in the solution space after unitarity cuts, we find that one can not let all $d^2$ terms be zero in $N_2$. 
This is an interesting feature that the CK-dual solution has a different $d$-dependence structure from that of Feynman diagrams, since all Feynman diagrams which can give contributions to topology of $N_2$ can not produce $d^2$ terms. 
The origin of this different structure is that here we require the integrand to not only pass unitarity cuts but also satisfy the requirement of CK duality.

\paragraph{On reduced tadpoles and massless bubbles.}

\begin{figure}[t]
	\centerline{\includegraphics[height=2.35cm]{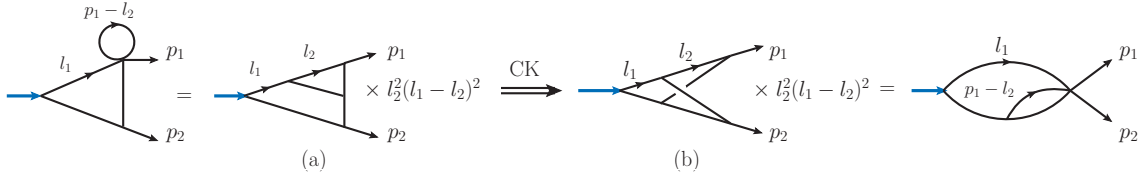}} 
	\caption{Via CK operation term in diagram (a) reducing to a tadpole may generate a term in diagram (b) reducing neither to tadpole nor massless bubble. } 
	\label{fig:tadpole_example}
\end{figure}

In the one-loop case, all terms with remaining free parameters can be reduced to massless bubbles or tadpoles, as shown in \eqref{eq:N11} in Section~\ref{sec:1loopsolution}.
One can exclude these terms when we construct the ansatz for the master numerator. 
We may wonder whether there is a similar feature at two loops, and this is the question that is also raised in Section~\ref{eq:2loopNumansatz}.
To check this, we collect all terms that will reduce to tadpoles or massless bubbles in the solution space of master numerators, and then we ask if they can be set to zero in the solution space. 
We find that for the first planar-ladder master topology, one can indeed make its numerator $N_1$ contain no tadpoles and massless bubbles in the solution space, but for the second master we are not able to do so. 
This implies that one can not drop all the terms corresponding to reduced tadpoles or massless bubbles from the starting ansatz. 
One way to understand this is that a tadpole or massless bubble term in one topology may become a term having non-trivial contribution in other topologies by CK operation. 
We show an example in Figure~\ref{fig:tadpole_example}.
Although such terms have no physical effect, they are needed in the numerators because of the requirement of CK duality.

\subsubsection{Constraints from tadpoles and massless bubbles}
\label{sec:constraintstadpole}

\begin{figure}[t]
	\centerline{\includegraphics[height=2.7cm]{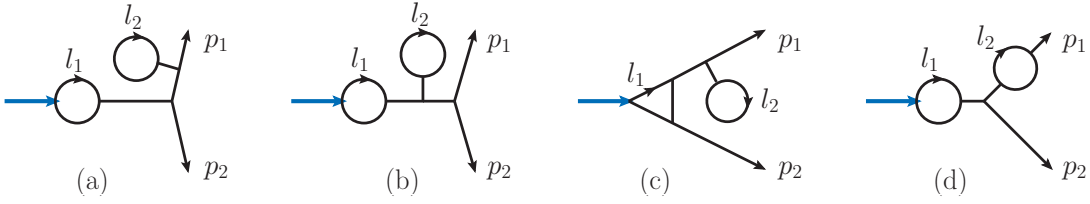}} 
	\caption{Tadpoles or massless bubbles which can be set to zero.
	} 
	\label{fig:tadpoles_zero}
\end{figure}

\begin{figure}[t]
	\centerline{\includegraphics[height=2.9cm]{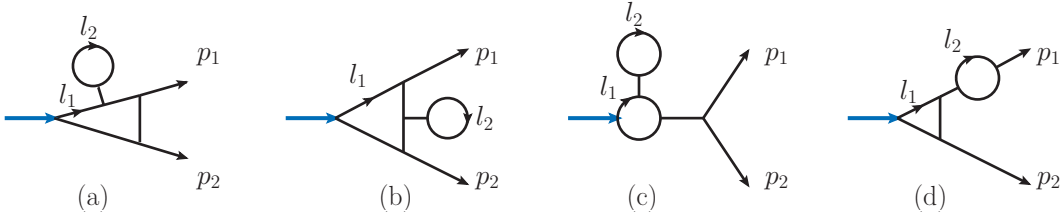}} 
	\caption{Tadpoles or massless bubbles which can \emph{not} be set to zero.} 
	\label{fig:tadpoles_nonzero}
\end{figure}

In this subsection, we consider the possible further constraints on the solution space by asking if one can set the numerators of trivalent topologies containing massless bubbles or tadpoles to be zero.
It is important to first emphasize that here the tadpoles or massless bubbles should \textit{not} be confused with the reduced tadpoles or massless bubbles discussed in the previous subsection.
In this subsection, we consider the trivalent topologies with a maximal number of propagators, and the tadpole or massless-bubble topologies all refer to the diagrams in Figure~\ref{fig:2looptadpoleandmassless}.
A complete set of CK relations will generate these topologies.
As we will discuss below, we find that some of them can be set to zero while others can not. They are shown in Figure~\ref{fig:tadpoles_zero} and Figure~\ref{fig:tadpoles_nonzero} respectively.

\paragraph{Constraints from tadpoles.}
We consider tadpole topologies first. 
We find that all the tadpoles 
on the external legs, \emph{i.e.}~(a) and (c) in Figure~\ref{fig:tadpoles_zero}, their numerators $N_{14}$ and $N_{12}$ can be set to zero. 
These tadpoles are related to massless bubbles via dual Jacobi relations. 
For tadpoles related to massive bubbles, however, we can not let all of them be zero but only $N_{15}$ for Figure~\ref{fig:tadpoles_zero}(b).
Requiring the above three tadpole numerators to be zero, we can solve for 91 parameters, leaving 144 parameters. 

\begin{figure}[t]
	\centerline{\includegraphics[height=2.5cm]{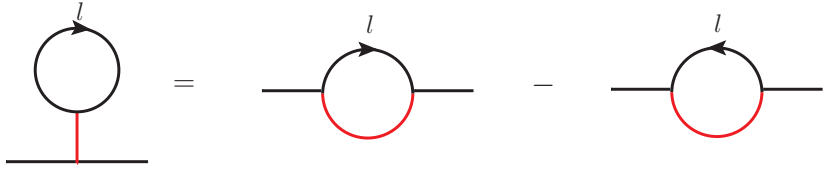}} 
	\caption{The generation of tadpoles from sub-bubble graphs using dual Jacobi relations. The red lines represent the propagators on which we apply CK operation.} 
	\label{fig:generate_tadpole}
\end{figure}

We provide an interpretation of why we may not set all tadpoles to be zero as follows. Through CK relations, the numerator of a tadpole topology is related to the numerator of topology containing a sub-bubble, such as shown in Figure~\ref{fig:generate_tadpole}. 
This means that the vanishing of a tadpole numerator will require the corresponding bubble diagram that generates it should only have even powers of the loop momentum flowing inside the bubble. 
This requirement can be too strong for a massive bubble since it also has to satisfy constraints directly from unitarity cuts. On the other hand, a massless bubble on the external leg is not constrained by unitarity cuts directly (although through CK relations it can be related to other topologies and is constrained by unitarity cuts indirectly). 
Thus the constraints are weaker in the latter case.

\paragraph{Constraints from massless bubbles.}

Next we consider topologies that contain a bubble on external legs. 
Although the massless bubble integral is scaleless and becomes zero after integration in dimensional regularization, there is still a subtlety at the integrand level.
For such trivalent topologies, there is an intermediate on-shell propagator which is divergent and makes the integrand apparently ill-defined. 

As pointed out in \cite{Bern:2013yya}, 
one can impose constraints such that the numerator has the same power-counting property as that of Feynman diagrams.
The numerator of a massless bubble obtained from Feynman rules satisfies the property that it contains two powers of the $l$ or $p_i$, where $l$ is the loop momentum flowing inside the massless bubble and $p_i$ is the corresponding external momentum. After integration, they will always be proportional to $p_i^2$ and cancel the on-shell propagator in the denominator, such as 
\begin{equation}
	\int d^{4-2\epsilon} l { \{l^2, l\cdot p_i \} \over l^2 (l-p_i)^2  p_i^2} \propto (p_i^2)^{-\epsilon} \,,
\end{equation}
which vanishes for $\epsilon<0$.
But the numerator constructed by CK duality in general may contain numerators not satisfying this property. 
For example, terms like  $(\varepsilon_1 \cdot \varepsilon_2)(p_1 \cdot p_2)^3$ would be ill-defined even after integration since the divergent propagator $1/p_i^2$ still exist. 

In our case, there are two bubble-on-external-leg topologies, Figure~\ref{fig:tadpoles_zero}(d) and Figure~\ref{fig:tadpoles_nonzero}(d).
For the first case, the whole numerator $N_{17}$ can be set to zero, which will solve for 40 parameters, leaving 104 parameters.
For the second topologies, one cannot set the full numerator $N_{13}$ to be zero.
We impose the condition that the terms which are not proportional to $l_2^2$, $(p_1-l_2)^2$ and $\varepsilon_1 \cdot l_2$ should be zero. Terms that proportional to $\varepsilon_1 \cdot l_2$ are allowed because after integration, $l_2$ will be replaced by $p_1$ and these terms will vanish because of the transverse condition $\varepsilon_1 \cdot p_1=0$. 
We find these conditions can indeed be satisfied for this topology. 
This can further solve for 22 parameters, leaving 82 free parameters in the solution space.

At this point, we further ask if we can let some other numerators be zero. We find that $N_9$ and $N_6$ can be set to zero in the above solution space. This will further solve for 8 and 9 parameters respectively, leaving 65 free parameters.
To summarize, in the solution space with 235 parameters, one can set $N_6, N_9, N_{12}, N_{14}, N_{15}$ and $N_{17}$ to be zero, and no other numerators can be set to zero.

\subsection{Origins of the remaining parameters}
After applying unitarity constraints and other conditions as discussed in previous subsections, 
we find the CK-dual solution space still contains a large number of free parameters.
All these parameters cancel in the final physical results. Below we discuss the origin of these parameters. 

The first origin of the parameters is related to the fact that the ansatz we construct includes terms that can be reduced to tadpoles and massless bubbles. Since the unitarity cut can not detect such terms, many of such parameters will remain undetermined. 
In the one-loop case, we have seen in Section~\ref{sec:1loopsolution} that all the remaining parameters in \eqref{eq:N11} are related to this type of origin. 
However, in the two-loop case, as discussed in Section~\ref{sec:2loopunitarity}, one can not exclude all terms that are reduced to tadpoles and massless bubbles, because of the requirement of CK duality.

\begin{figure}[t]
	\centerline{\includegraphics[height=2.cm]{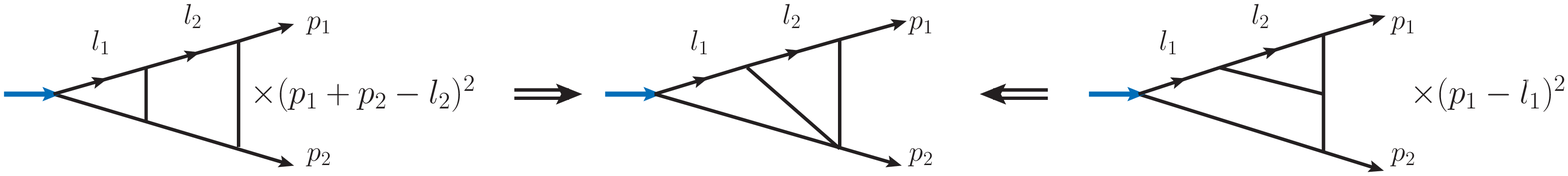}} 
	\caption{Different trivalent topologies may reduce to the same sub-topology.} 
	\label{fig:redistribution}
\end{figure}

The second origin is more non-trivial and it comes from the freedom of redistributing terms in different topologies. In the CK-dual ansatz, all the integrals correspond to trivalent topologies which have a maximal number of propagators. If a term in the numerator is proportional to one of the propagators, this term can be reduced to a sub-topology. It is possible that different trivalent topologies may reduce to the same sub-topology and such an example is shown in Figure~\ref{fig:redistribution}. 
In other words, the same contribution of a certain sub-topology may be expressed in terms of different maximal topologies, and this fact leads to some degrees of freedom when expressing the integrand.

We have checked that most of the free parameters are associated with the second origin. Actually, only two parameters (out of 235) are purely due to the first trivial origin, which reduces only to tadpole or massless bubble sub-topologies. Below we briefly comment on the physical meaning of the free parameters.
Since all parameters cancel in the final result, they should be understood as ``gauge" parameters in the sense that different choices of the parameters give different but physically equivalent representations of the same form factor. Such kind of freedom generally exists for loop integrands of form factors or amplitudes. 
The real non-trivial point here is that we also require the integrand solution to satisfy the CK duality. In particular, different solutions in the solution space preserve all dual Jacobi relations. The free parameters in this sense can be understood as CK-preserving deformation parameters. 
We mention that similar solution space with a large number of parameters was also observed for the three-point half-BPS form factors in ${\cal N}=4$ SYM up to four-loops \cite{Lin:2021kht,Lin:2021qol, Lin:2021lqo}.
As discussed in \cite{Lin:2021kht, Lin:2021qol}, the CK-preserving deformation is also related to the generalized gauge transformation associated with the operator insertion.  
The existence of such a deformation is due to the color-singlet nature of the operator and thus is a special feature for form factors. We refer interested reader to  \cite{Lin:2021kht, Lin:2021qol} for more details on this point.

The two master numerators in the solution space of  235 free parameters are explicitly given in Appendix~\ref{app:2loopnum} where the free parameters are set to zero for simplicity. 
The full numerators with all free parameters are provided in the auxiliary files.

\section{Integral reduction and integration} 
\label{sec:integration}
In this section we will show that all the parameters disappear after integral reduction. 
We also discuss the integrated results and focus on differences between the results of the $d$-dimensional integrand and $4$-dimensional integrand.
Since the one-loop case is relatively simple, we will mainly focus on the two-loop case.

To simplify the integrand, we first evaluate the color factors. This can be done by first expanding the structure constants in terms of color trace products using \eqref{def:StrucCons} and then applying the contraction rule:
\begin{equation}
	\sum_{a=1}^{N_c^2-1} (T^a)_i^{~j} (T^a)_k^{~l} = \delta_i^{~l} \delta_k^{~j} - {1\over N_c} \delta_i^{~j} \delta_k^{~l} \,.
\end{equation} 
For the two-loop case, after contracting internal color indices, all the color factors will either be proportional to $N_c^2 {\rm tr}(T^{a_1} T^{a_2}) $ or equal to zero. We find that only the first 5 topologies in Figure \ref{fig:2looptopo} have non-zero color factors, so all other topologies will not contribute to the full form factor in gauge theory and hence we will omit them in the following discussion.\footnote{The topologies with zero color factors can have a non-trivial contribution in the study of double copy for gravitational quantities. We will not consider this in the present work.} 
We define the five non-trivial color factors as $C_i = N_c^2 {\rm tr}(T^{a_1} T^{a_2}) \times c_i$ and $c_i$ are given as
\begin{equation}\label{eq:generalSudakovFF}
	c_1=4, \quad c_2=2, \quad c_3=2, \quad c_4=4, \quad c_5=4 \,.
\end{equation}
For completeness, we also give their symmetry factors coming from the isomorphism of graphs:
\begin{equation}\label{eq:generalSudakovFF}
	S_1=2, \quad S_2=1, \quad S_3=4, \quad S_4=2, \quad S_5=4 \,.
\end{equation}
The full integrand of the two-loop form factor can be written as:
\begin{equation}\label{eq:generalSudakovFF}
	\hat{\mathcal{F}}_2{}^{(2)}=i^{2}g^4 N_c^2 {\rm tr}(T^{a_1} T^{a_2}) {\cal F}_2^{(2)}, \qquad {\cal F}_2^{(2)} = \sum _{\sigma _2} \sum _{i=1}^5 \int \prod_{j=1}^2 \frac{d^dl_j}{(2\pi)^d} \frac{1}{S_i}\frac{c_iN_i}{\
		\Pi_aD_{i,a}} \,.
\end{equation}

\subsection{Integral reduction}
In this subsection, we do the integral reduction and check that the remaining parameters all cancel in the final results.  
Notice that our integrand contains terms like $\varepsilon_i \cdot l_j$.
A common strategy would be to first reduce such terms by the PV reduction \cite{Passarino:1978jh} and then perform integration-by-part (IBP) reduction \cite{Chetyrkin:1981qh, Tkachov:1981wb}. 
Here we will use an alternative way based on gauge-invariant basis projection.

Since the final result must be gauge invariant, one can expand the final result by a set of gauge-invariant basis. For a general discussion about gauge-invariant basis, one may refer to Section~2 of \cite{Boels:2017gyc}. For two-point form factor there is only one gauge-invariant basis, which is defined as:
\begin{equation}\label{eq:gauge_basis}
	B_0=(\varepsilon_1 \cdot \varepsilon_2)(p_1 \cdot p_2)-(\varepsilon_1 \cdot p_2)(\varepsilon_2 \cdot p_1) \,,
\end{equation}
and this is obviously equivalent to the tree-level Sudakov form factor in \eqref{eq:tree-sudakovFF}.
The integrand can be expanded as:
\begin{equation}\label{eq:gauge_basis_expand}
	{\cal F}_2^{(l)}(\varepsilon_i,p_j,l_k)=B_0\; f(p_j,l_k) \,,
\end{equation}
where
$f(p_j,l_k)$ contains only Lorentz product momentum variables and we can perform IBP directly. To get the expression of $f(p_j,l_k)$ we multiply $B_0$ to both sides of \eqref{eq:gauge_basis_expand} and sum over the polarization vectors by the rule in \eqref{eq:contraction_rule}, which gives
\begin{equation}
	\sum _{\text{helicities}} B_0 \, {\cal F}_2^{(l)}(\varepsilon_i,p_j,l_k) =f(p_j,l_k)\sum _{\text{helicities}} B_0^2=f(p_j,l_k)(d-2)(p_1 \cdot p_2)^2 \,.
\end{equation}
Here we point out a technical subtlety: since the integrand is not manifestly gauge invariant,  the reference momentum $\xi$ introduced through the helicity sum \eqref{eq:contraction_rule} will not vanish in $f(p_j,l_k)$. We take $\xi$ as another external momentum and perform the IBP with a set of propagator basis including $\xi$ (here we use the $\mathtt{LiteRed}$ package \cite{Lee:2013mka}). We find that the $\xi$ dependence indeed disappears after IBP, which also provides a cross-check for the result.

After collecting the coefficients of master integrals, we find the parameters cancel and the coefficients we obtain are consistent with the known result \cite{Gehrmann:2005pd}. 
For the convenience of discussing the integrated results, we also use that the following form
\begin{equation}
	\begin{split}
		\hat{\cal F}^{(l)}_{2} 
		=&\tilde{g}^{2l}  {\rm tr}(T^{a_1} T^{a_2})  \, {\cal F}^{(0)}_{2} {\cal I}^{(l)},
	\end{split}
\end{equation}
where
\begin{equation}
	\tilde{g}^{2}=g^2 \frac{(4\pi e^{- \gamma})^\epsilon N_c}{16\pi^2} \,,
\end{equation}
and the expansions of ${\cal I}^{(l = 1,2)}$ in terms of master integrals are given in Appendix~\ref{masterint}.

\subsection{Integrated results}

The integrand numerators we obtain are polynomials of spacetime dimension parameter $d=4-2\epsilon$.
The term depending on $\epsilon$ will not be fixable by four-dimensional cuts. 
For one-loop amplitudes, such terms will contribute to rational terms after integration. 
One may wonder what is the contribution of the $\epsilon$-dependent terms for the Sudakov form factor.
In this subsection we address this problem by considering the integrated results.

Considering first the one-loop result, we separate the result into two terms
\begin{equation}
	{\cal I}^{(1)} = {\cal I}_{d=4}^{(1)} + {\cal I}_{\epsilon}^{(1)} \,,
\end{equation}
where ${\cal I}_{d=4}^{(1)}$ is obtained by setting $d=4$ in the full integrand, and ${\cal I}_{\epsilon}^{(1)}$ is from the remaining integrand that are linear in $\epsilon$.
After integration, they are
\begin{equation}
	\begin{split}
		&{\cal I}_{d=4}^{(1)}=-\frac{2}{\epsilon ^2}+\frac{\pi ^2}{6}+\frac{14 \zeta (3) \epsilon }{3}+\frac{47 \pi ^4 \epsilon ^2}{720}+O\left(\epsilon ^3\right) , \\
		&{\cal I}_{\epsilon}^{(1)}=-2 \epsilon -6 \epsilon ^2+\frac{1}{6} \left(\pi ^2-84\right) \epsilon ^3+O\left(\epsilon ^4\right) .
	\end{split}
\end{equation}
We can see that ${\cal I}_{\epsilon}^{(1)}$ 
starts from $\epsilon$-order and will not affect the divergent and finite terms.

Next for the two-loop result, we separate the integrand into three terms
\begin{equation}
	{\cal I}^{(2)} = {\cal I}_{d=4}^{(2)} + {\cal I}_{\epsilon^1}^{(2)} + {\cal I}_{\epsilon^2}^{(2)} \,,
\end{equation}
where ${\cal I}_{d=4}^{(2)}$ is obtained by setting $d=4$ in the full integrand, and ${\cal I}_{\epsilon^n}^{(2)}$ are from the integrand contribution that is linear in $\epsilon^n$.
After integration, they are
\begin{align}
		&{\cal I}_{d=4}^{(2)}=\frac{4}{\epsilon ^4}-\frac{11}{3 \epsilon ^3}+\frac{-\frac{64}{9}-\frac{\pi ^2}{3}}{\epsilon ^2}+\frac{-\frac{50 \zeta (3)}{3}+\frac{11 \pi ^2}{6}-\frac{56}{27}}{\epsilon }+\left(\frac{22 \zeta (3)}{9}-\frac{7 \pi ^4}{30}+\frac{32 \pi ^2}{9}+\frac{5045}{81}\right)+O\left(\epsilon ^1\right), \nonumber
		\\
		&{\cal I}_{\epsilon^1}^{(2)}=-\frac{1}{3 \epsilon ^2}+\frac{64}{9 \epsilon }+\left(\frac{272}{27}+\frac{\pi ^2}{6}\right)+\left(\frac{2 \zeta (3)}{9}-\frac{8 \pi ^2}{9}-\frac{4397}{81}\right) \epsilon +O\left(\epsilon ^2\right), \nonumber
		\\
		&{\cal I}_{\epsilon^2}^{(2)}=-8 \epsilon -40 \epsilon ^2+\left(\frac{4 \pi ^2}{3}-136\right) \epsilon ^3+O\left(\epsilon ^4\right) .
\end{align}
Unlike the one-loop case, now we can see that terms proportional to $\epsilon$ in the integrand will provide contribution starting from $1/\epsilon^2$ divergence. This shows clearly that $d$-dimensional cuts are necessary to get the correct form factor result.

\section{Summary and discussion}
\label{sec:discussion}

In this paper, we study the color-kinematics duality for one and two-loop Sudakov form factors of ${\rm tr}(F^2)$ in the non-supersymmetric pure YM theory.
We find that massless bubble and tadpole type topologies are necessarily involved. 
Although they are zero after integration, they are required to preserve the CK duality. 
The CK-dual numerators have the same powers of loop momenta as expected from Feynman diagrams. 
Interestingly, despite the non-trivial constraints from both unitarity cuts and dual Jacobi relations, the two-loop solution space still contains a large number of free parameters. 
Our results imply that the CK duality may also hold in the pure YM theory at three loops or for higher-point form factors, which are certainly interesting to be explored further.

Since the color-kinematics duality plays an important role in constructing gravitational amplitudes via double copy, it would be interesting to explore the double-copy of the form factors. 
For the Sudakov form factor studied in this paper, one may naively apply double copy. However, to have a consistent gravitational quantity, it is crucial to ensure that the double-copy quantity not only preserves the diffeomorphism invariance but is also consistent with all unitarity cuts in the gravitational theory. 
Some progress has been made recently for the double copy of tree-level form factors in \cite{Lin:2021pne} where it was found that certain spurious poles become physical poles after the double copy. The new tree-level double-copy results also provide important building blocks for the further study of unitarity cuts at the loop level.
It would be highly interesting to study further the double copy for loop form factors which we leave for future works.

\section*{Acknowledgements}
It is a pleasure to thank Guanda Lin and Siyuan Zhang for discussions. 
This work is supported in part by the National Natural Science Foundation of China (Grants No.~12175291, 11935013, 11822508, 12047503),
and by the Key Research Program of the Chinese Academy of Sciences, Grant NO. XDPB15.
We also thank the support of the HPC Cluster of ITP-CAS.

\appendix

\section{Complete two-loop dual Jacobi relations}\label{app:CKrelation}
Here are all CK relations we have between 17 cubic graphs in Figure~\ref{fig:2looptopo} and Figure~\ref{fig:2looptadpoleandmassless}. 
\eqref{eq:A1}--\eqref{eq:A7} can generate other seven numerators detectable by cuts. 
\eqref{eq:A8}--\eqref{eq:A15} are relations to generate numerators of graphs with a massless bubble or tadpole.  
\eqref{eq:A16}--\eqref{eq:A23} are other CK relations. They will not provide further constraints for the 15 numerators obtained from \eqref{eq:A1} to~\eqref{eq:A15}.
\begin{align}
	N_3&=N_1-N_2\label{eq:A1}\,,\\
	N_4&=N_2+N_2 \left[l_1,l_1-l_2,p_1,p_2\right]\,,\\
	N_5&=N_2\left[l_1,p_1-l_2,p_1,p_2\right]+N_2\left[l_1,l_1+l_2,p_1,p_2\right]\,,\\
	N_6&=N_1-N_1 \left[p_1+p_2-l_1,l_2,p_1,p_2\right]\,,\\
	N_7&=N_1-N_1 \left[l_1,l_2,p_2,p_1\right]\,,\\
	N_8&=N_7+N_7 \left[l_1,l_1-l_2,p_1,p_2\right]\,,\\
	N_9&=N_7+N_7 \left[l_1,p_1+p_2-l_2,p_1,p_2\right]\label{eq:A7}\,,\\
	N_{10}&=N_4-N_4\left[l_1,-l_2,p_1,p_2\right]\label{eq:A8}\,,\\
	N_{11}&=-N_5+N_5\left[l_1,-l_2,p_1,p_2\right]\,,\\
	N_{12}&=-N_{13}+N_{13}\left[l_1,-l_2,p_1,p_2\right]\,,\\
	N_{13}&=N_2+N_2\left[l_1,p_1-l_2,p_1,p_2\right]\,,\\
	N_{14}&=N_{17}-N_{17}\left[l_1,-l_2,p_1,p_2\right]\,,\\
	N_{15}&=N_9-N_9\left[l_1,-l_2,p_1,p_2\right]\,,\\
	N_{16}&=N_8-N_8\left[l_1,-l_2,p_1,p_2\right]\,,\\
	N_{17}&=N_6+N_6\left[l_1,p_1-l_2,p_1,p_2\right]\label{eq:A15}\,,\\
	N_8&=N_4-N_4 \left[l_1,l_2,p_2,p_1\right]\label{eq:A16}\,,\\
	N_9&=N_6-N_6 \left[l_1,l_2,p_2,p_1\right]\,,\\
	N_{11}&=N_{12}+N_{10}\,,\\
	N_{13}&=N_{17}+N_{13}\left[p_1+p_2-l_1,l_2,p_1,p_2\right]\,,\\
	N_{14}&=-N_{12}+N_{12}\left[p_1+p_2-l_1,l_2,p_1,p_2\right]\,,\\
	N_{15}&=N_{16}-N_{16}\left[p_1+p_2-l_1,l_2,p_1,p_2\right]\,,\\
	N_{15}&=N_{14}-N_{14}\left[l_1,l_2,p_2,p_1\right]\,,\\
	N_{16}&=N_{10}-N_{10}\left[l_1,l_2,p_2,p_1\right]\label{eq:A23}\,.
\end{align}

\section{Two-loop master numerators}
\label{app:2loopnum}
In this appendix we provide the two-loop master numerators after using the unitarity constraints in Section~\ref{sec:2loopunitarity}, which contain 235 unfixed parameters. Here we set all free parameters to zero for simplicity, and the full numerators with parameters are provided in the auxiliary files.
{\small
	
	\begin{dmath*}
		N_1 =
		2 \mathcal{E} s^3-(d-4) \mathcal{E} D_1 s^2-\frac{1}{16} \left(d^2+20 d-108\right) \mathcal{E} D_2 s^2+2 (d-4) \mathcal{E} D_3 s^2+\frac{1}{8} \left(d^2+4 d-44\right) \mathcal{E} D_4 s^2-(d-4) \mathcal{E} D_5 s^2-\frac{1}{16} \left(d^2+20 d-108\right) \mathcal{E} D_6 s^2-4 \mathcal{K}_{12} \mathcal{K}_{21} s^2+4 (d-4) \mathcal{K}_{14} \mathcal{K}_{21} s^2-4 (d-4) \mathcal{K}_{13} \mathcal{K}_{23} s^2+4 (d-4) \mathcal{K}_{14} \mathcal{K}_{23} s^2+4 (d-4) \mathcal{K}_{13} \mathcal{K}_{24} s^2-8 (d-4) \mathcal{K}_{14} \mathcal{K}_{24} s^2+\frac{1}{4} (d-2)^2 \mathcal{E} D_2^2 s+\frac{1}{4} (d-2)^2 \mathcal{E} D_4^2 s+\frac{1}{4} (d-2)^2 \mathcal{E} D_6^2 s+2 (d-4) \mathcal{E} D_7^2 s+\frac{1}{4} (d-6) (d-2) \mathcal{E} D_1 D_2 s-\frac{1}{4} (d-4) (d-2) \mathcal{E} D_2 D_3 s-\frac{1}{4} (d-8) (d-2) \mathcal{E} D_1 D_4 s-\frac{1}{2} (d-3) (d-2) \mathcal{E} D_2 D_4 s+\frac{1}{2} (d-4) (d-2) \mathcal{E} D_3 D_4 s+2 (d-4) \mathcal{E} D_1 D_5 s-\frac{1}{2} (d-2) \mathcal{E} D_2 D_5 s-\frac{1}{4} (d-8) (d-2) \mathcal{E} D_4 D_5 s-\frac{1}{2} (d-2) \mathcal{E} D_1 D_6 s+\frac{1}{4} (d-6) (d-2) \mathcal{E} D_2 D_6 s-\frac{1}{4} (d-4) (d-2) \mathcal{E} D_3 D_6 s-\frac{1}{2} (d-3) (d-2) \mathcal{E} D_4 D_6 s+\frac{1}{4} (d-6) (d-2) \mathcal{E} D_5 D_6 s-2 (d-4) \mathcal{E} D_1 D_7 s+(d-2) \mathcal{E} D_2 D_7 s-2 (d-2) \mathcal{E} D_4 D_7 s-2 (d-4) \mathcal{E} D_5 D_7 s+(d-2) \mathcal{E} D_6 D_7 s+2 (d-4) D_1 \mathcal{K}_{12} \mathcal{K}_{21} s+\frac{1}{2} (d-4) (d+2) D_2 \mathcal{K}_{12} \mathcal{K}_{21} s-4 (d-4) D_3 \mathcal{K}_{12} \mathcal{K}_{21} s+2 (d-4) D_5 \mathcal{K}_{12} \mathcal{K}_{21} s-\frac{1}{2} (d-6) (d-4) D_6 \mathcal{K}_{12} \mathcal{K}_{21} s-4 (d-4) D_1+4 (d-2) D_2 \mathcal{K}_{13} \mathcal{K}_{21} s-8 (d-2) D_4 \mathcal{K}_{13} \mathcal{K}_{21} s
	\end{dmath*}

	\begin{dmath*}
		\mid(\textrm{equation continued}) \\
		+4 (d-2) D_6 \mathcal{K}_{13} \mathcal{K}_{21} s+4 (d-4) D_7 \mathcal{K}_{13} \mathcal{K}_{21} s-(d-6) (d-4) D_1 \mathcal{K}_{14} \mathcal{K}_{21} s-2 (d-2)^2 D_2 \mathcal{K}_{14} \mathcal{K}_{21} s+2 (d-4) (d-2) D_3 \mathcal{K}_{14} \mathcal{K}_{21} s+(d-2) (d+4) D_4 \mathcal{K}_{14} \mathcal{K}_{21} s-(d-4) (d-2) D_5 \mathcal{K}_{14} \mathcal{K}_{21} s-(d-2) d D_6 \mathcal{K}_{14} \mathcal{K}_{21} s-4 (d-4) D_7 \mathcal{K}_{14} \mathcal{K}_{21} s-4 (d-2) D_2 \mathcal{K}_{12} \mathcal{K}_{23} s \mathcal{K}_{13}+8 (d-2) D_4 \mathcal{K}_{12} \mathcal{K}_{23} s+4 (d-4) D_5 \mathcal{K}_{12} \mathcal{K}_{23} s-4 (d-2) D_6 \mathcal{K}_{12} \mathcal{K}_{23} s-4 (d-4) D_7 \mathcal{K}_{12} \mathcal{K}_{23} s+4 (d-2) D_2 \mathcal{K}_{12} \mathcal{K}_{24} s-(d-2) (d+4) D_4 \mathcal{K}_{12} \mathcal{K}_{24} s-4 (d-4) D_5 \mathcal{K}_{12} \mathcal{K}_{24} s \mathcal{K}_{21} s+(d-2) d D_6 \mathcal{K}_{12} \mathcal{K}_{24} s+4 (d-4) D_7 \mathcal{K}_{12} \mathcal{K}_{24} s+(d-4) (d-2) D_1 \mathcal{K}_{14} \mathcal{K}_{24} s+(d-4) (d-2) D_2 \mathcal{K}_{14} \mathcal{K}_{24} s-2 (d-4) (d-2) D_3 \mathcal{K}_{14} \mathcal{K}_{24} s+(d-4) (d-2) D_5 \mathcal{K}_{14} \mathcal{K}_{24} s+(d-4) (d-2) D_6 \mathcal{K}_{14} \mathcal{K}_{24} s-\frac{1}{2} (d-2) \mathcal{E} D_1 D_6^2-\frac{1}{2} (d-2) \mathcal{E} D_1 D_2 D_4-\frac{1}{2} (d-2) \mathcal{E} d_2^2 D_5+\frac{1}{2} (d-2) \mathcal{E} D_2 D_4 D_5+\frac{1}{2} (d-2) \mathcal{E} D_1 D_2 D_6+\frac{1}{2} (d-2) \mathcal{E} D_1 D_4 D_6+\frac{1}{2} (d-2) \mathcal{E} D_2 D_5 D_6-\frac{1}{2} (d-2) \mathcal{E} D_4 D_5 D_6+\frac{1}{2} (d-2) \mathcal{E} D_2^2 D_7+\frac{1}{2} (d-2) \mathcal{E} D_6^2 D_7-(d-2) \mathcal{E} D_2 D_6 D_7+2 (d-2) D_2^2 \mathcal{K}_{13} \mathcal{K}_{21}+2 (d-2) D_6^2 \mathcal{K}_{13} \mathcal{K}_{21}-4 (d-2) D_2 D_6 \mathcal{K}_{13} \mathcal{K}_{21}-2 (d-2) D_2^2 \mathcal{K}_{14} \mathcal{K}_{21}-2 (d-2) D_6^2 \mathcal{K}_{14} \mathcal{K}_{21}+4 (d-2) D_2 D_6 \mathcal{K}_{14} \mathcal{K}_{21}-2 (d-2) D_2^2 \mathcal{K}_{12} \mathcal{K}_{23}-2 (d-2) D_6^2 \mathcal{K}_{12} \mathcal{K}_{23}+4 (d-2) D_2 D_6 \mathcal{K}_{12} \mathcal{K}_{23}+2 (d-2) D_2^2 \mathcal{K}_{12} \mathcal{K}_{24}+2 (d-2) D_6^2 \mathcal{K}_{12} \mathcal{K}_{24}-4 (d-2) D_2 D_6 \mathcal{K}_{12} \mathcal{K}_{24} \,.
	\end{dmath*}

	\begin{dmath*}
		N_2=
		-(d-2)^2 \mathcal{E} D_4^3-\frac{1}{4} (d-2) (3 d-10) s \mathcal{E} D_4^2-\frac{1}{4} (d-2) (d+2) \mathcal{E} D_1 D_4^2+\frac{1}{2} (d-2)^2 \mathcal{E} D_2 D_4^2-(d-2)^2 \mathcal{E} D_3 D_4^2+\frac{1}{4} \left(-13 d^2+76 d-148\right) \mathcal{E} D_5 D_4^2+\frac{3}{2} (d-2)^2 \mathcal{E} D_6 D_4^2+\frac{3}{2} (d-2)^2 \mathcal{E} D_7 D_4^2+(d-2) d \mathcal{K}_{13} \mathcal{K}_{21} D_4^2-2 (d-2) d \mathcal{K}_{14} \mathcal{K}_{21} D_4^2-(d-2) d \mathcal{K}_{12} \mathcal{K}_{23} D_4^2+2 (d-2) d \mathcal{K}_{12} \mathcal{K}_{24} D_4^2+\frac{1}{12} \left(-9 d^2+90 d-248\right) \mathcal{E} D_1^2 D_4+\frac{3}{2} (d-6) (d-4) \mathcal{E} D_2^2 D_4+\frac{1}{6} \left(-9 d^2+48 d-80\right) \mathcal{E} D_5^2 D_4-\frac{1}{2} (d-2)^2 \mathcal{E} D_6^2 D_4-\frac{1}{2} (d-2)^2 \mathcal{E} D_7^2 D_4+\frac{1}{4} \left(d^2-18 d+64\right) s^2 \mathcal{E} D_4+\frac{1}{12} \left(3 d^2-18 d-16\right) s \mathcal{E} D_1 D_4+\frac{1}{4} (d-10) (d-2) s \mathcal{E} D_2 D_4+\frac{1}{4} (d-2) (d+6) \mathcal{E} D_1 D_2 D_4+\frac{1}{4} \left(-5 d^2+48 d-124\right) s \mathcal{E} D_3 D_4+\frac{1}{2} \left(3 d^2-28 d+68\right) \mathcal{E} D_1 D_3 D_4+3 (3 d-10) \mathcal{E} D_2 D_3 D_4+\frac{1}{12} \left(-21 d^2+144 d-164\right) s \mathcal{E} D_5 D_4-\frac{1}{4} (d-2) (d+4) \mathcal{E} D_1 D_5 D_4+\frac{3}{4} \left(3 d^2-20 d+44\right) \mathcal{E} D_2 D_5 D_4-2 \left(d^2-8 d+18\right) \mathcal{E} D_3 D_5 D_4+\frac{1}{2} (d-2)^2 s \mathcal{E} D_6 D_4+\frac{1}{4} (d-2) (d+2) \mathcal{E} D_1 D_6 D_4+\left(-2 d^2+17 d-38\right) \mathcal{E} D_2 D_6 D_4+\left(2 d^2-17 d+38\right) \mathcal{E} D_3 D_6 D_4+\frac{1}{4} \left(11 d^2-72 d+148\right) \mathcal{E} D_5 D_6 D_4+\frac{1}{2} (d-2) (2 d-11) s \mathcal{E} D_7 D_4+\frac{1}{4} (d-2) (d+4) \mathcal{E} D_1 D_7 D_4-\frac{1}{2} (d-2) d \mathcal{E} D_2 D_7 D_4+\frac{1}{2} (d-2)^2 \mathcal{E} D_3 D_7 D_4+\frac{1}{4} \left(11 d^2-70 d+144\right) \mathcal{E} D_5 D_7 D_4-(d-3) (d-2) \mathcal{E} D_6 D_7 D_4-(d-6) (d-4) s \mathcal{K}_{12} \mathcal{K}_{21} D_4+\frac{4}{3} \left(3 d^2-30 d+88\right) D_1 \mathcal{K}_{12} \mathcal{K}_{21} D_4-(d-2) d D_2 \mathcal{K}_{12} \mathcal{K}_{21} D_4-(d-2) d D_3 \mathcal{K}_{12} \mathcal{K}_{21} D_4-32 D_5 \mathcal{K}_{12} \mathcal{K}_{21} D_4+\left(d^2-10 d+8\right) s \mathcal{K}_{13} \mathcal{K}_{21} D_4-2 \left(3 d^2-30 d+80\right) D_1 \mathcal{K}_{13} \mathcal{K}_{21} D_4-(d-2) (d+4) D_2 \mathcal{K}_{13} \mathcal{K}_{21} D_4+2 (d-2) (d+1) D_3 \mathcal{K}_{13} \mathcal{K}_{21} D_4
	\end{dmath*}

	\begin{dmath*}
		\mid(\textrm{equation continued}) \\
		+\frac{1}{3} \left(9 d^2-66 d+224\right) D_5 \mathcal{K}_{13} \mathcal{K}_{21} D_4-(d-2)^2 D_6 \mathcal{K}_{13} \mathcal{K}_{21} D_4-(d-2) d D_7 \mathcal{K}_{13} \mathcal{K}_{21} D_4-(d-2)^2 s \mathcal{K}_{14} \mathcal{K}_{21} D_4-2 (d-2) D_1 \mathcal{K}_{14} \mathcal{K}_{21} D_4+2 (d-2) (d+2) D_2 \mathcal{K}_{14} \mathcal{K}_{21} D_4-4 (d-2) (d+1) D_3 \mathcal{K}_{14} \mathcal{K}_{21} D_4-2 (d-2) (d+1) D_5 \mathcal{K}_{14} \mathcal{K}_{21} D_4+2 (d-2) d D_6 \mathcal{K}_{14} \mathcal{K}_{21} D_4+2 (d-2) d D_7 \mathcal{K}_{14} \mathcal{K}_{21} D_4-(d-6) (d-4) s \mathcal{K}_{12} \mathcal{K}_{23} D_4-2 \left(3 d^2-28 d+76\right) D_1 \mathcal{K}_{12} \mathcal{K}_{23} D_4-(d-2) d D_2 \mathcal{K}_{12} \mathcal{K}_{23} D_4+2 (d-2) D_3 \mathcal{K}_{12} \mathcal{K}_{23} D_4+\frac{1}{3} \left(3 d^2-54 d+224\right) D_5 \mathcal{K}_{12} \mathcal{K}_{23} D_4+(d-2)^2 D_6 \mathcal{K}_{12} \mathcal{K}_{23} D_4+(d-2) d D_7 \mathcal{K}_{12} \mathcal{K}_{23} D_4+(d-2) d s \mathcal{K}_{13} \mathcal{K}_{23} D_4+\frac{8}{3} \left(3 d^2-30 d+88\right) D_1 \mathcal{K}_{13} \mathcal{K}_{23} D_4+4 (d-2) D_2 \mathcal{K}_{13} \mathcal{K}_{23} D_4-\frac{8}{3} \left(3 d^2-30 d+88\right) D_5 \mathcal{K}_{13} \mathcal{K}_{23} D_4-4 (d-2) D_6 \mathcal{K}_{13} \mathcal{K}_{23} D_4-(d-2) d s \mathcal{K}_{14} \mathcal{K}_{23} D_4-4 (d-2) D_2 \mathcal{K}_{14} \mathcal{K}_{23} D_4+4 (d-2) D_6 \mathcal{K}_{14} \mathcal{K}_{23} D_4+(d-2)^2 s \mathcal{K}_{12} \mathcal{K}_{24} D_4+2 (d-2) D_1 \mathcal{K}_{12} \mathcal{K}_{24} D_4+2 (d-2) (d+2) D_2 \mathcal{K}_{12} \mathcal{K}_{24} D_4-4 (d-2) D_3 \mathcal{K}_{12} \mathcal{K}_{24} D_4+2 (d-2) (d+1) D_5 \mathcal{K}_{12} \mathcal{K}_{24} D_4-2 (d-2) d D_6 \mathcal{K}_{12} \mathcal{K}_{24} D_4-2 (d-2) d D_7 \mathcal{K}_{12} \mathcal{K}_{24} D_4-(d-2) d s \mathcal{K}_{13} \mathcal{K}_{24} D_4-4 (d-2) D_2 \mathcal{K}_{13} \mathcal{K}_{24} D_4+4 (d-2) D_6 \mathcal{K}_{13} \mathcal{K}_{24} D_4+2 (d-2)^2 s \mathcal{K}_{14} \mathcal{K}_{24} D_4+4 (d-2) D_1 \mathcal{K}_{14} \mathcal{K}_{24} D_4-4 (d-2) D_5 \mathcal{K}_{14} \mathcal{K}_{24} D_4+\frac{3}{4} (d-6) (d-4) \mathcal{E} D_5^3+\frac{1}{6} (3 d-22) s \mathcal{E} D_1^2+\frac{1}{12} \left(-9 d^2+78 d-128\right) s \mathcal{E} D_5^2+\frac{3}{4} (d-6) (d-4) \mathcal{E} D_1 D_5^2+\frac{1}{12} \left(21 d^2-162 d+352\right) \mathcal{E} D_2 D_5^2+\frac{3}{2} (d-6) (d-4) \mathcal{E} D_2 D_6^2+(d-6) s \mathcal{E} D_7^2
		%
		%
		+\frac{3}{4} (d-6) (d-4) \mathcal{E} D_1 D_7^2+\frac{1}{2} (d-2) \mathcal{E} D_2 D_7^2+\frac{3}{4} (d-6) (d-4) \mathcal{E} D_5 D_7^2+\frac{1}{2} (2-d) \mathcal{E} D_6 D_7^2+\frac{1}{6} (9 d-46) s^2 \mathcal{E} D_1+\frac{3}{4} (d-6) (d-4) \mathcal{E} D_1^2 D_2+\frac{1}{8} \left(-d^2-20 d+108\right) s^2 \mathcal{E} D_2+\frac{1}{4} \left(-d^2+12 d-12\right) s \mathcal{E} D_1 D_2+\left(d^2-13 d+34\right) \mathcal{E} D_2^2 D_3+\frac{1}{8} (d-18) (d-6) s^2 \mathcal{E} D_3+\frac{1}{2} (d-2) s \mathcal{E} D_1 D_3-\frac{1}{2} (d-2) d s \mathcal{E} D_2 D_3-\frac{3}{2} (d-6) (d-4) \mathcal{E} D_1 D_2 D_3+\frac{3}{2} (d-6) (d-4) \mathcal{E} D_2^2 D_5+\frac{1}{48} \left(-3 d^2+12 d-44\right) s^2 \mathcal{E} D_5+\frac{1}{6} (9 d-74) s \mathcal{E} D_1 D_5+\frac{1}{12} \left(3 d^2-36 d+44\right) s \mathcal{E} D_2 D_5+\frac{1}{6} \left(6 d^2-39 d+74\right) \mathcal{E} D_1 D_2 D_5+2 \left(d^2-8 d+18\right) \mathcal{E} D_2 D_3 D_5+\frac{8}{3} \mathcal{E} D_1^2 D_6-\frac{3}{2} (d-6) (d-4) \mathcal{E} D_2^2 D_6-\frac{3}{4} (d-6) (d-4) \mathcal{E} D_5^2 D_6+\frac{1}{3} (10-3 d) s \mathcal{E} D_1 D_6+(d-2) s \mathcal{E} D_2 D_6-\frac{1}{4} (d-2) (d+6) \mathcal{E} D_1 D_2 D_6-\frac{1}{4} (d-6) (d-2) s \mathcal{E} D_3 D_6+(2-d) \mathcal{E} D_1 D_3 D_6+\left(-2 d^2+17 d-38\right) \mathcal{E} D_2 D_3 D_6+\frac{1}{4} \left(5 d^2-36 d+44\right) s \mathcal{E} D_5 D_6+\frac{1}{12} \left(-9 d^2+84 d-172\right) \mathcal{E} D_1 D_5 D_6+\frac{1}{4} \left(-13 d^2+116 d-276\right) \mathcal{E} D_2 D_5 D_6-\frac{3}{2} (d-6) (d-4) \mathcal{E} D_2^2 D_7-\frac{3}{2} (d-6) (d-4) \mathcal{E} D_5^2 D_7+\frac{1}{16} \left(d^2+20 d-172\right) s^2 \mathcal{E} D_7+\frac{1}{2} (34-5 d) s \mathcal{E} D_1 D_7+\frac{1}{2} \left(-d^2+12 d-28\right) s \mathcal{E} D_2 D_7+\frac{1}{6} \left(-6 d^2+39 d-74\right) \mathcal{E} D_1 D_2 D_7+\frac{1}{2} (2-d) s \mathcal{E} D_3 D_7-\frac{1}{2} (d-2) d \mathcal{E} D_2 D_3 D_7+\frac{1}{12} \left(9 d^2-84 d+188\right) s \mathcal{E} D_5 D_7-\frac{3}{2} (d-6) (d-4) \mathcal{E} D_1 D_5 D_7+\frac{1}{6} \left(-15 d^2+123 d-278\right) \mathcal{E} D_2 D_5 D_7-\frac{1}{4} (d-6) (d+2) s \mathcal{E} D_6 D_7+\frac{1}{12} \left(9 d^2-84 d+172\right) \mathcal{E} D_1 D_6 D_7+\frac{3}{2} (d-6) (d-4) \mathcal{E} D_2 D_6 D_7+(d-2) \mathcal{E} D_3 D_6 D_7+\frac{1}{12} \left(9 d^2-84 d+172\right) \mathcal{E} D_5 D_6 D_7
	\end{dmath*}

	\begin{dmath*}
		\mid(\textrm{equation continued}) \\
		-2 (d-2) D_2^2 \mathcal{K}_{12} \mathcal{K}_{21}-16 D_5^2 \mathcal{K}_{12} \mathcal{K}_{21}+\frac{1}{3} \left(3 d^2-36 d+136\right) s D_1 \mathcal{K}_{12} \mathcal{K}_{21}+(d-4) (d+6) s D_2 \mathcal{K}_{12} \mathcal{K}_{21}-\frac{2}{3} \left(3 d^2-33 d+94\right) D_1 D_2 \mathcal{K}_{12} \mathcal{K}_{21}+2 (3 d-10) s D_3 \mathcal{K}_{12} \mathcal{K}_{21}+(d-2) (d+2) D_2 D_3 \mathcal{K}_{12} \mathcal{K}_{21}+\frac{1}{3} \left(-9 d^2+60 d-136\right) s D_5 \mathcal{K}_{12} \mathcal{K}_{21}+\frac{2}{3} \left(3 d^2-30 d+88\right) D_1 D_5 \mathcal{K}_{12} \mathcal{K}_{21}+\left(20-d^2\right) D_2 D_5 \mathcal{K}_{12} \mathcal{K}_{21}-\frac{2}{3} \left(3 d^2-30 d+88\right) D_1 D_6 \mathcal{K}_{12} \mathcal{K}_{21}+(d-2) (d+2) D_2 D_6 \mathcal{K}_{12} \mathcal{K}_{21}-2 (d-2) D_3 D_6 \mathcal{K}_{12} \mathcal{K}_{21}+16 D_5 D_6 \mathcal{K}_{12} \mathcal{K}_{21}+\frac{1}{2} (d-6) (d-4) s D_7 \mathcal{K}_{12} \mathcal{K}_{21}-\frac{2}{3} \left(3 d^2-30 d+88\right) D_1 D_7 \mathcal{K}_{12} \mathcal{K}_{21}+(d-2) d D_2 D_7 \mathcal{K}_{12} \mathcal{K}_{21}+16 D_5 D_7 \mathcal{K}_{12} \mathcal{K}_{21}-8 s^2 \mathcal{K}_{13} \mathcal{K}_{21}+2 (d-2) D_2^2 \mathcal{K}_{13} \mathcal{K}_{21}+16 D_5^2 \mathcal{K}_{13} \mathcal{K}_{21}+\frac{1}{4} \left(-9 d^2+58 d-136\right) s D_1 \mathcal{K}_{13} \mathcal{K}_{21}-(d-8) (d-2) s D_2 \mathcal{K}_{13} \mathcal{K}_{21}+\frac{2}{3} \left(3 d^2-30 d+88\right) D_1 D_2 \mathcal{K}_{13} \mathcal{K}_{21}-2 (d-2) s D_3 \mathcal{K}_{13} \mathcal{K}_{21}-2 (d-2) (d+3) D_2 D_3 \mathcal{K}_{13} \mathcal{K}_{21}+\frac{1}{12} \left(15 d^2-54 d+152\right) s D_5 \mathcal{K}_{13} \mathcal{K}_{21}-\frac{4}{3} \left(3 d^2-30 d+76\right) D_1 D_5 \mathcal{K}_{13} \mathcal{K}_{21}+\frac{1}{3} \left(-9 d^2+54 d-152\right) D_2 D_5 \mathcal{K}_{13} \mathcal{K}_{21}+8 s D_6 \mathcal{K}_{13} \mathcal{K}_{21}+\frac{4}{3} \left(3 d^2-30 d+76\right) D_1 D_6 \mathcal{K}_{13} \mathcal{K}_{21}+(d-2) d D_2 D_6 \mathcal{K}_{13} \mathcal{K}_{21}+4 (d-2) D_3 D_6 \mathcal{K}_{13} \mathcal{K}_{21}+2 (d-10) D_5 D_6 \mathcal{K}_{13} \mathcal{K}_{21}-8 s D_7 \mathcal{K}_{13} \mathcal{K}_{21}+\frac{4}{3} \left(3 d^2-30 d+76\right) D_1 D_7 \mathcal{K}_{13} \mathcal{K}_{21}
		%
		+(d-2) (d+4) D_2 D_7 \mathcal{K}_{13} \mathcal{K}_{21}-16 D_5 D_7 \mathcal{K}_{13} \mathcal{K}_{21}-2 (d-2) D_6 D_7 \mathcal{K}_{13} \mathcal{K}_{21}+8 s^2 \mathcal{K}_{14} \mathcal{K}_{21}+\frac{1}{3} \left(-3 d^2+30 d-88\right) D_1^2 \mathcal{K}_{14} \mathcal{K}_{21}-4 (d-2) D_2^2 \mathcal{K}_{14} \mathcal{K}_{21}+\frac{1}{6} \left(15 d^2-54 d+88\right) s D_1 \mathcal{K}_{14} \mathcal{K}_{21}-(d-4) (d-2) s D_2 \mathcal{K}_{14} \mathcal{K}_{21}+2 (d-2) D_1 D_2 \mathcal{K}_{14} \mathcal{K}_{21}+(d-6) (d-2) s D_3 \mathcal{K}_{14} \mathcal{K}_{21}+4 (d-2) (d+1) D_2 D_3 \mathcal{K}_{14} \mathcal{K}_{21}+\left(-5 d^2+26 d-64\right) s D_5 \mathcal{K}_{14} \mathcal{K}_{21}+\frac{1}{3} \left(3 d^2-42 d+112\right) D_1 D_5 \mathcal{K}_{14} \mathcal{K}_{21}+2 (d-2) (d+2) D_2 D_5 \mathcal{K}_{14} \mathcal{K}_{21}+4 (d-2) D_1 D_6 \mathcal{K}_{14} \mathcal{K}_{21}-2 (d-2) d D_2 D_6 \mathcal{K}_{14} \mathcal{K}_{21}-2 (d-2) D_5 D_6 \mathcal{K}_{14} \mathcal{K}_{21}+\left(d^2-10 d+40\right) s D_7 \mathcal{K}_{14} \mathcal{K}_{21}+\frac{1}{3} \left(3 d^2-18 d+64\right) D_1 D_7 \mathcal{K}_{14} \mathcal{K}_{21}-2 (d-2) (d+3) D_2 D_7 \mathcal{K}_{14} \mathcal{K}_{21}+\frac{1}{3} \left(-3 d^2+30 d-88\right) D_5 D_7 \mathcal{K}_{14} \mathcal{K}_{21}+2 (d-2) D_6 D_7 \mathcal{K}_{14} \mathcal{K}_{21}+2 (d-2) D_2^2 \mathcal{K}_{12} \mathcal{K}_{23}+\frac{1}{3} \left(3 d^2-30 d+88\right) D_5^2 \mathcal{K}_{12} \mathcal{K}_{23}+\frac{1}{12} \left(3 d^2+66 d-296\right) s D_1 \mathcal{K}_{12} \mathcal{K}_{23}-4 (d-4) s D_2 \mathcal{K}_{12} \mathcal{K}_{23}+\frac{4}{3} \left(3 d^2-30 d+76\right) D_1 D_2 \mathcal{K}_{12} \mathcal{K}_{23}+2 (d-2) s D_3 \mathcal{K}_{12} \mathcal{K}_{23}+2 (d-2) D_2 D_3 \mathcal{K}_{12} \mathcal{K}_{23}+\frac{1}{4} \left(-3 d^2-18 d+40\right) s D_5 \mathcal{K}_{12} \mathcal{K}_{23}+\frac{1}{3} \left(-3 d^2+30 d-88\right) D_1 D_5 \mathcal{K}_{12} \mathcal{K}_{23}+\left(-d^2+6 d-24\right) D_2 D_5 \mathcal{K}_{12} \mathcal{K}_{23}+\frac{2}{3} \left(3 d^2-30 d+88\right) D_1 D_6 \mathcal{K}_{12} \mathcal{K}_{23}+(d-2) d D_2 D_6 \mathcal{K}_{12} \mathcal{K}_{23}-4 (d-2) D_3 D_6 \mathcal{K}_{12} \mathcal{K}_{23}-\frac{2}{3} \left(3 d^2-27 d+82\right) D_5 D_6 \mathcal{K}_{12} \mathcal{K}_{23}+4 (d-2) s D_7 \mathcal{K}_{12} \mathcal{K}_{23}+\frac{1}{3} \left(3 d^2-30 d+88\right) D_1 D_7 \mathcal{K}_{12} \mathcal{K}_{23}+(d-2) d D_2 D_7 \mathcal{K}_{12} \mathcal{K}_{23}+\frac{1}{3} \left(-3 d^2+30 d-88\right) D_5 D_7 \mathcal{K}_{12} \mathcal{K}_{23}
	\end{dmath*}

	\begin{dmath*}
		\mid(\textrm{equation continued}) \\
		+2 (d-2) D_6 D_7 \mathcal{K}_{12} \mathcal{K}_{23}-4 (d-2) D_2^2 \mathcal{K}_{13} \mathcal{K}_{23}-\frac{2}{3} \left(3 d^2-30 d+88\right) D_5^2 \mathcal{K}_{13} \mathcal{K}_{23}+\frac{1}{2} \left(3 d^2-30 d+104\right) s D_1 \mathcal{K}_{13} \mathcal{K}_{23}+(d-4) (d-2) s D_2 \mathcal{K}_{13} \mathcal{K}_{23}-\frac{4}{3} \left(3 d^2-30 d+88\right) D_1 D_2 \mathcal{K}_{13} \mathcal{K}_{23}+\frac{1}{6} \left(-3 d^2+30 d-40\right) s D_5 \mathcal{K}_{13} \mathcal{K}_{23}+\frac{2}{3} \left(3 d^2-30 d+88\right) D_1 D_5 \mathcal{K}_{13} \mathcal{K}_{23}+\frac{4}{3} \left(3 d^2-30 d+88\right) D_2 D_5 \mathcal{K}_{13} \mathcal{K}_{23}-\frac{4}{3} \left(3 d^2-30 d+88\right) D_1 D_6 \mathcal{K}_{13} \mathcal{K}_{23}+4 (d-2) D_2 D_6 \mathcal{K}_{13} \mathcal{K}_{23}+\frac{4}{3} \left(3 d^2-30 d+88\right) D_5 D_6 \mathcal{K}_{13} \mathcal{K}_{23}-\frac{2}{3} \left(3 d^2-30 d+88\right) D_1 D_7 \mathcal{K}_{13} \mathcal{K}_{23}+\frac{2}{3} \left(3 d^2-30 d+88\right) D_5 D_7 \mathcal{K}_{13} \mathcal{K}_{23}-8 s^2 \mathcal{K}_{14} \mathcal{K}_{23}-\frac{2}{3} \left(3 d^2-30 d+88\right) D_5^2 \mathcal{K}_{14} \mathcal{K}_{23}-4 (d-2) D_6^2 \mathcal{K}_{14} \mathcal{K}_{23}-\frac{1}{2} (d-14) (d-4) s D_1 \mathcal{K}_{14} \mathcal{K}_{23}-(d-2) d s D_2 \mathcal{K}_{14} \mathcal{K}_{23}+4 (d-2) D_1 D_2 \mathcal{K}_{14} \mathcal{K}_{23}+\frac{1}{6} \left(9 d^2-90 d+232\right) s D_5 \mathcal{K}_{14} \mathcal{K}_{23}+\frac{2}{3} \left(3 d^2-30 d+88\right) D_1 D_5 \mathcal{K}_{14} \mathcal{K}_{23}-4 (d-2) D_2 D_5 \mathcal{K}_{14} \mathcal{K}_{23}+4 (d-2) s D_6 \mathcal{K}_{14} \mathcal{K}_{23}-4 (d-2) D_1 D_6 \mathcal{K}_{14} \mathcal{K}_{23}+4 (d-2) D_2 D_6 \mathcal{K}_{14} \mathcal{K}_{23}+4 (d-2) D_5 D_6 \mathcal{K}_{14} \mathcal{K}_{23}-\frac{2}{3} \left(3 d^2-30 d+88\right) D_1 D_7 \mathcal{K}_{14} \mathcal{K}_{23}+\frac{2}{3} \left(3 d^2-30 d+88\right) D_5 D_7 \mathcal{K}_{14} \mathcal{K}_{23}+\frac{1}{3} \left(3 d^2-30 d+88\right) D_5^2 \mathcal{K}_{12} \mathcal{K}_{24}+\frac{1}{6} \left(-15 d^2+78 d-200\right) s D_1 \mathcal{K}_{12} \mathcal{K}_{24}-2 (d-2) D_1 D_2 \mathcal{K}_{12} \mathcal{K}_{24}-2 (d-2) s D_3 \mathcal{K}_{12} \mathcal{K}_{24}+\frac{1}{3} \left(9 d^2-30 d+56\right) s D_5 \mathcal{K}_{12} \mathcal{K}_{24}
		%
		+\frac{1}{3} \left(-3 d^2+30 d-88\right) D_1 D_5 \mathcal{K}_{12} \mathcal{K}_{24}+2 (d-2) d D_2 D_5 \mathcal{K}_{12} \mathcal{K}_{24}-2 (d-2) (d+2) D_2 D_6 \mathcal{K}_{12} \mathcal{K}_{24}+4 (d-2) D_3 D_6 \mathcal{K}_{12} \mathcal{K}_{24}+2 (d-2) D_5 D_6 \mathcal{K}_{12} \mathcal{K}_{24}-(d-4) (d+2) s D_7 \mathcal{K}_{12} \mathcal{K}_{24}+\frac{1}{3} \left(3 d^2-30 d+88\right) D_1 D_7 \mathcal{K}_{12} \mathcal{K}_{24}-2 (d-2) (d+1) D_2 D_7 \mathcal{K}_{12} \mathcal{K}_{24}+\frac{1}{3} \left(-3 d^2+30 d-88\right) D_5 D_7 \mathcal{K}_{12} \mathcal{K}_{24}-2 (d-2) D_6 D_7 \mathcal{K}_{12} \mathcal{K}_{24}+8 s^2 \mathcal{K}_{13} \mathcal{K}_{24}+4 (d-2) D_2^2 \mathcal{K}_{13} \mathcal{K}_{24}-\frac{2}{3} \left(3 d^2-30 d+88\right) D_5^2 \mathcal{K}_{13} \mathcal{K}_{24}+\frac{1}{6} \left(-3 d^2+30 d-136\right) s D_1 \mathcal{K}_{13} \mathcal{K}_{24}-(d-4) (d-2) s D_2 \mathcal{K}_{13} \mathcal{K}_{24}+\frac{3}{2} (d-6) (d-4) s D_5 \mathcal{K}_{13} \mathcal{K}_{24}+\frac{2}{3} \left(3 d^2-30 d+88\right) D_1 D_5 \mathcal{K}_{13} \mathcal{K}_{24}-4 (d-2) D_2 D_6 \mathcal{K}_{13} \mathcal{K}_{24}-\frac{2}{3} \left(3 d^2-30 d+88\right) D_1 D_7 \mathcal{K}_{13} \mathcal{K}_{24}+\frac{2}{3} \left(3 d^2-30 d+88\right) D_5 D_7 \mathcal{K}_{13} \mathcal{K}_{24}+\frac{2}{3} \left(3 d^2-30 d+88\right) D_1^2 \mathcal{K}_{14} \mathcal{K}_{24}+\frac{2}{3} \left(3 d^2-36 d+100\right) D_5^2 \mathcal{K}_{14} \mathcal{K}_{24}+2 (d-2)^2 s D_2 \mathcal{K}_{14} \mathcal{K}_{24}-4 (d-2) D_1 D_2 \mathcal{K}_{14} \mathcal{K}_{24}+4 (d-2) s D_3 \mathcal{K}_{14} \mathcal{K}_{24}+(d-2) d s D_5 \mathcal{K}_{14} \mathcal{K}_{24}-\frac{4}{3} \left(3 d^2-33 d+94\right) D_1 D_5 \mathcal{K}_{14} \mathcal{K}_{24}-4 (d-2) s D_6 \mathcal{K}_{14} \mathcal{K}_{24}+4 (d-2) D_5 D_6 \mathcal{K}_{14} \mathcal{K}_{24}-(d-4) (d-2) s D_7 \mathcal{K}_{14} \mathcal{K}_{24}-4 (d-2) D_1 D_7 \mathcal{K}_{14} \mathcal{K}_{24}+4 (d-2) D_2 D_7 \mathcal{K}_{14} \mathcal{K}_{24}+4 (d-2) D_5 D_7 \mathcal{K}_{14} \mathcal{K}_{24}-4 (d-2) D_6 D_7 \mathcal{K}_{14} \mathcal{K}_{24} \,.
	\end{dmath*}
	
}

\section{Master integral}\label{masterint}
\label{app:masterintegral}

The definition of ${\cal I}^{(l)}$ is:
\begin{equation}
{\cal I}^{(l)} =(e^{\epsilon \gamma})^{l} \sum _{\sigma _2} \sum _{\Gamma_i} \int \prod_{j=1}^l \frac{d^dl_j}{i\pi^{d/2}} \frac{1}{S_i}\frac{c_iN_i}{\
		\Pi_aD_{i,a}} \,.
\end{equation}
After IBP, we can expand ${\cal I}^{(l)}$ by master integrals:
\begin{equation}
	{\cal I}^{(l)} =(-s)^{- \epsilon l}S_R^l\;(\sum _{\rm i}C_{\rm i}^{(l)}I_{\rm i}^{(l)}),
\end{equation}
where $S_R=\frac{e^{\epsilon\gamma}}{\Gamma\left(1-\epsilon\right)}$ is an overall factor and $I_{\rm i}^{(l)}$ means the $i$th master integral of $l$-loop case. 
Topologies of masters are shown in Figure \ref{fig:master_integral_graph} and their explicit results are given below.
\begin{figure}[t]
	\centerline{\includegraphics[height=2.3cm]{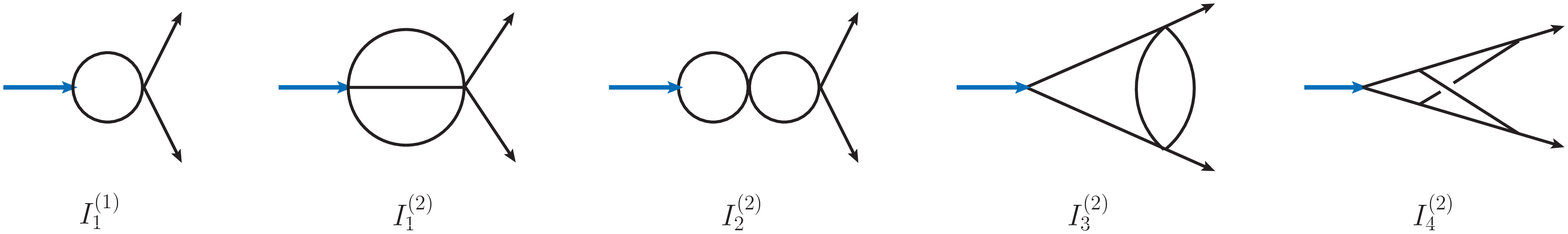}} 
	\caption{Graphs corresponding to master integrals. } 
	\label{fig:master_integral_graph}
\end{figure}

\subsection{One-loop}
Master integral for one-loop is 
\begin{equation}
		I_{\rm 1}^{(1)}= \frac{\Gamma(1-\epsilon)^3\Gamma(\epsilon+1)}{\epsilon\Gamma(2-2\epsilon )}
		=\frac{1}{\epsilon }+2+4\epsilon+(8-2\zeta(3))\epsilon^2+\left(-4 \zeta (3)+16-\frac{\pi ^4}{30}\right) \epsilon^3+O\left(\epsilon^4\right)\,.
\end{equation}
Coefficient of this master is
\begin{equation}
	C_{\rm 1}^{(1)}=\frac{4}{d-4}-\frac{4}{d-2}+10-d\,.
\end{equation}

\subsection{Two-loop}
Master integrals for two-loop are below. Notice that the previous three integrals have exact expressions in terms of Gamma functions. Integral $I_4$ can only be expressed in hypergeometric functions and we do not list it. For details you can refer to Section 2 of~\cite{Gehrmann:2005pd}.
\begin{align}
	I_{\rm 1}^{(2)}=&-\frac{\Gamma[1-\epsilon]^5\Gamma[1+2 \epsilon]}{2\epsilon(-1+2\epsilon)\Gamma[3-3\epsilon]}\notag\\
	=&-\frac{1}{4\epsilon}-\frac{13}{8}-\frac{115\epsilon}{16}+\left(\frac{5\zeta(3)}{2}-\frac{865}{32}\right)\epsilon^2+\left(\frac{65\zeta(3)}{4}-\frac{5971}{64}+\frac{\pi^4}{24}\right)\epsilon^3+O\left(\epsilon^4\right)\,,\\
	I_{\rm 2}^{(2)}=&\frac{\Gamma[1-\epsilon]^6\Gamma[1+\epsilon^2]}{\epsilon ^2 \Gamma[2-2\epsilon]^2}\notag\\
	=&\frac{1}{\epsilon^2}+\frac{4}{\epsilon}+12-4 (\zeta(3)-8)\epsilon+\left(-16\zeta(3)+80-\frac{\pi^4}{15}\right)\epsilon^2\notag\\
	&-\frac{4}{15}\left(180\zeta (3)+45\zeta(5)-720+\pi^4\right)\epsilon^3+O\left(\epsilon^4\right)\,,\\
	I_{\rm 3}^{(2)}=&\frac{\Gamma[1-2\epsilon]\Gamma[1-\epsilon]^4\Gamma[1+\epsilon]\Gamma[1+2\epsilon]}{2\epsilon^2(-1+2\epsilon)\Gamma[2-3\epsilon]}\notag\\
	=&\frac{1}{2\epsilon^2}+\frac{5}{2\epsilon}+\frac{1}{6}\left(57+\pi^2\right)+\left(\frac{5}{6}\left(39+\pi^2\right)-4\zeta(3)\right)\epsilon\notag\\
	&+\frac{1}{30}\left(-600\zeta(3)+3165+95\pi^2-\pi^4\right)\epsilon^2\notag\\
	&+\frac{1}{6}\left(-8\left(57+\pi^2\right)\zeta(3)-144\zeta(5)+1995+65\pi^2-\pi^4\right)\epsilon^3+O\left(\epsilon^4\right)\,,\\
	I_{\rm 4}^{(2)}=&\frac{1}{\epsilon^4}-\frac{5\pi^2}{6\epsilon^2}-\frac{27\zeta(3)}{\epsilon}-\frac{23\pi^4}{36}+\left(8\pi^2\zeta(3)-117\zeta(5)\right)\epsilon+\left(267\zeta(3)^2-\frac{19\pi^6}{315}\right)\epsilon^2\notag\\
	&+\left(\frac{109\pi^4\zeta(3)}{10}+40\pi^2\zeta(5)+6\zeta(7)\right)\epsilon^3+O\left(\epsilon^4\right)\,.
\end{align}

Coefficients of these masters for the two-loop form factor are
\begin{equation}
	\begin{split}
		C_{\rm 1}^{(2)}=&24d-\frac{1175}{3(d-4)}-\frac{2}{d-3}+\frac{96}{d-2}+\frac{116}{9(d-1)}+\frac{525}{16(2d-7)}\\
		&+\frac{107}{144(2d-5)}-\frac{1388}{3(d-4)^2}-\frac{32}{(d-2)^2}-\frac{192}{(d-4)^3}-\frac{1955}{8}\,,\\
		C_{\rm 2}^{(2)}=&d^2-20d+\frac{32}{d-4}-\frac{48}{d-2}+\frac{16}{(d-4)^2}+\frac{16}{(d-2)^2}+100\,,\\
		C_{\rm 3}^{(2)}=&\frac{27d}{2}+\frac{103}{3(d-4)}+\frac{80}{d-2}+\frac{10}{3(d-1)}+\frac{75}{16(2d-7)}\\
		&+\frac{119}{48(2d-5)}+\frac{24}{(d-4)^2}-\frac{32}{(d-2)^2}-\frac{609}{8}\,,\\
		C_{\rm 4}^{(2)}=&\frac{3(d-3)(3 d-8)}{4(2d-7)(2d-5)}\,.
	\end{split}
\end{equation}


\providecommand{\href}[2]{#2}\begingroup\raggedright\endgroup

\end{document}